\documentclass[preprint]{aastex}

\begin{document}

\title{
The precession of orbital plane and the significant variabilities
of binary pulsars}

\author{Biping Gong}

\affil{ Department of Astronomy, Nanjing University, Nanjing
210093,PR.China}

\email{bpgong@nju.edu.cn}

\begin{abstract}
 There are two kinds of expressions on the precession of
orbital plane of a binary pulsar system, which are given by Barker
$\&$ O'Connell (1975) and Apostolatos et al. (1994), Kidder (1995)
respectively. This paper points out that these two kinds of
orbital precession velocities are actually obtained by the same
Lagrangian under different degrees of freedom. Correspondingly the
former expression is not consistent with the conservation of the
total angular momentum vector; whereas the latter one is. Damour
$\&$ Sch\"afer (1988) and  Wex $\&$ Kopeikin (1999) have applied
Barker $\&$ O'Connell's orbital precession velocity in pulsar
timing measurement. This paper applies Apostolatos et al. $\&$
Kidder's orbital precession velocity in pulsar timing measurement.
We analyze that Damour $\&$ Sch\"afer's treatment corresponds to
negligible Spin-Orbit induced precession of periastron. Whereas
the effects corresponding to Wex $\&$ Kopeikin and this paper are
both significant (however they are not equivalent). The
observational data of two typical binary pulsars, PSR~J2051-0827
and PSR~J1713+0747 apparently support significant Spin-Orbit
coupling effect. Further more, the discrepancies between Wex $\&$
Kopeikin and this paper can be tested on specific binary pulsars
with orbital plane nearly edge on. If the orbital period
derivative of double-pulsar system PSRs~J0737-3039~A and B, with
orbital inclination angle $i=87.7_{-29}^{+17}$deg, is much larger
than that of the gravitational radiation induced one, then the
expression of this paper is supported, otherwise Wex $\&$
Kopeikin's expression is supported.
\end{abstract}


\maketitle

\section{Introduction}
In the gravitational two-body problem with spin, each body is
precessing in the gravitational field of its companion, with
precession velocity of 1 Post-Newtonian order (PN) (Barker $\&$
O'Connell 1975, hereafter BO). This precession velocity is widely
accepted. But how the orbital plane reacts to the torque caused by
the precession of the two bodies has two kinds of treatments. BO's
orbital precession velocity is obtained by assuming that the
angular momentum vector, ${\bf L}$, precesses at the same velocity
as the Runge-Lenz vector, ${\bf A}$.

On the other hand, in the study of  the modulation of the
gravitational wave by Spin-Orbit (S-L) coupling effect on merging
binaries (Apostolatos et al. 1994, Kidder~1995, hereafter AK)
obtain an orbital precession velocity that satisfies  the
conservation of the total angular momentum, ${\bf J}$, and the
triangle constraint, ${\bf J}={\bf L}+{\bf S}$ (where ${\bf S}$ is
the sum of spin angular momenta of two bodies, ${\bf S}_1$ and
${\bf S}_2$).

The discrepancy in the assumptions leads to different behaviors in
physics between BO and AK's orbital precession velocities. The
former isn't consistent with the triangle constraint, whereas the
latter is. And the reason of this discrepancy is that the former
actually assumes that the four vectors, ${\bf S}_1$, ${\bf S}_2$,
${\bf r}_1$ and  ${\bf r}_2$ (${\bf r}_1$, ${\bf r}_2$ are
position vectors of the two bodies respectively), are independent;
whereas, the latter  assumes that the independent vectors are
either ${\bf S}_1$, ${\bf r}_1$ and ${\bf r}_2$; or ${\bf S}_2$,
${\bf r}_1$ and ${\bf r}_2$.

The discrepancy between BO and AK's precession velocities is
analogous  to the following case. The motion of a small mass at
the bottom of a clock pendulum can be described in $x-y$ plane.
However if we treat the degree of freedom of this small mass as 2,
then this small mass can move every in the 2-dimension space, and
the length of the pendulum is not a constant. In other words, once
the length of the pendulum is a constant, the degree of freedom is
1 instead of 2. Correspondingly if the free vectors of a binary
system is 4, as treated by BO, then the triangle constraint cannot
be satisfied (or ${\bf J}$ cannot be a constant vector). On the
other hand, if the triangle constraint is satisfied, the number of
free vectors is 3, instead of 4.

 In the application to pulsar timing
measurement, BO's orbital precession velocity is treated in two
different ways, by Damour $\&$ Sch\"afer (1988) and  Wex $\&$
Kopeikin (1999) respectively. The former one predicts
insignificant  S-L coupling induced precession of periastron,
$\dot{\omega}^S$, and the  derivative of orbital period,
$\dot{P}_b$; whereas the latter predicts significant
$\dot{\omega}^S$ and  $\dot{P}_b$. The paper points out that the
discrepancy is due to  Damour $\&$ Sch\"afer  and Wex $\&$
Kopeikin calculated effects in different kinds of coordinate
systems, the former coordinate system is not at rest to "an
observer" at the Solar System Baryon center (SSB), whereas, the
latter is at rest to it. In other words, the former coordinate
system has non-zero acceleration to SSB; whereas, the latter one
has zero acceleration to SSB.

This paper calculates observational effect corresponding to AK's
orbital precession velocity, which uses the same coordinate system
as that of Wex $\&$ Kopeikin (1999). Significant $\dot{\omega}^S$
and $\dot{P}_b$ are given, but they are not equivalent to the
results  given by Wex $\&$ Kopeikin (1999). The validity of these
two expressions can be tested by specific binary pulsars with
orbital inclination close to $\pi/2$, i.e., the double-pulsar
system PSRs~J0737-3039~A and B.

This paper contains four parts: (a) the physical discrepancy
between BO and AK's orbital precession velocity (Sect~2,3); (b)
the derivation of S-L coupling induced effect corresponding to
AK's expression of orbital precession (Sect~4,5,9); (c) the
discrepancy between the coordinate systems used by Damour $\&$
Sch\"afer (1988) and Wex $\&$ Kopeikin (1999), as well as the
relationships among three kinds of S-L coupling induced effects,
Damour $\&$ Sch\"afer and Wex $\&$ Kopeikin and this paper
(Sect~6); (d) confrontation of the three S-L coupling effects with
observational data of PSR~J2051-0827 and PSR~J1713+0747, and
different predictions on  $\dot{\omega}^S$ and $\dot{P}_b$ of
PSRs~J0737-3039~A and B  by  different S-L coupling  models
(Sect~7,8).

\section{Orbital precession velocity}
This section introduces the derivation of the orbital precession
velocity of BO and AK.
\subsection{Derivation of BO's orbital
precession velocity} BO's two-body equation was the first
gravitational two-body equation with spin, which consists of two
parts, the precession velocity of the spin angular momenta vectors
of body one and body two, and the precession velocity of the
orbital angular momentum vector. Body one precesses in the
gravitational field of body two, with precession velocity (BO),
\begin{equation}\label{e1x1}
 \dot{\bf \Omega}_{1}= \frac {L
(4+3m_{2}/m_{1})}{2r^{3}}\,\hat{\bf L}+\frac {S_{2}}{2r^{3}}
\left[\hat{\bf S}_2-3(\hat{\bf L}\cdot\hat{\bf S}_2)\,\hat{\bf
L}\right]
\end{equation}
where $\hat{\bf L}$, $\hat{\bf S}_1 $  and $\hat{\bf S}_2$ are
unit vectors of the orbital angular momentum, the spin angular
momentum of star 1 and star 2, respectively.

$\dot{\bf \Omega}_{2}$ can be obtained by exchanging the subscript
1 and 2 at the right side of Eq.($\ref{e1x1}$). The first term  of
Eq.($\ref{e1x1}$) represents the geodetic (de Sitter) precession,
which corresponds to the precession of ${\bf S}_{1}$ around ${\bf
L}$, it is 1PN due to $\frac{L}{r^3}\sim
(\frac{v}{c})^2(\frac{v}{r})$; and the second term represents the
Lense-Thirring precession, ${\bf S}_{1}$ around ${\bf S}_{2}$,
which is  $\frac{S}{L}$ times smaller than the first, therefore,
it corresponds to 1.5PN. The the precession velocity of the spin
angular momenta vectors is confirmed by other authors using
different method.

However for the precession velocity of the orbit, there are
different expressions. BO's orbital precession velocity is given
as follows.

The total Hamiltonian for the gravitational two-body problem with
spin is given (BO,  Damour $\&$ Sch\"afer~1988),
\begin{equation}
\label{e1.2}H=H_{N}+H_{1PN}+H_{2PN}+H_{S} \ ,
\end{equation}
where $H_{N}$, $H_{1PN}$ and $H_{2PN}$ are the Newtonian, the
first and  second order post-Newtonian terms respectively. $H_{S}$
is the spin-orbit interaction Hamiltonian (BO, Damour $\&$
Sch\"afer~1988),
\begin{equation}
\label{e1.2hs} H_{S}=\sum\limits_{\alpha=1}^{2}
(2+3\frac{m_{\alpha+1}}{m_{\alpha}})(\frac{{\bf
S}_{\alpha}\cdot{\bf L}}{r^{3}}) \ ,
\end{equation}
where $\alpha+1$ is meant modulo 2 (2+1=1), $ m_{1}$, $m_{2}$ are
the masses of the two stars, respectively,  $r=a(1-e^{2})^{1/2}$,
$a$ is the semi-major axis, $e$ is the eccentricity of the orbit.
Notice we uses $G=c=1$ until discussing observational effects in
Sec~5 to Sec~7.


The BO equation describes the secular effect of the orbital plane
by a rotational velocity vector, $\dot{\Omega}_S$, acting on some
instantaneous Newtonian ellipse. Damour $\&$ Sch\"afer (1988)
computed $\dot{\Omega}_S$ in a simple manner by making full use of
the Hamiltonian method. The functions of the canonically conjugate
phase space variables ${\bf r}$ and ${\bf p}$ are defined as
\begin{equation}
\label{e1.5} {\bf L}({\bf r},{\bf p})={\bf r}\times {\bf p} \ ,
\end{equation}
\begin{equation}
\label{e1.6} {\bf A}({\bf r},{\bf p})={\bf p}\times {\bf
L}-GM\mu^{2}\frac{{\bf r}}{r} \ ,
\end{equation}
where ${\bf r}={\bf r}_1-{\bf r}_2$, $M=m_1+m_2$, $\mu=m_1m_2/M$.
The vector ${\bf A}$ is the Runge-Lenz vector (first discovered by
Lagrange). The instantaneous Newtonian ellipse evolves according
to the fundamental equations of Hamiltonian dynamics (Damour $\&$
Sch\"afer 1988)
\begin{equation}
\label{e1.7} \dot{\bf L}=\{{\bf L},H\} \ ,
\end{equation}
\begin{equation}
\label{e1.8} \dot{\bf A}=\{{\bf A},H\} \ ,
\end{equation}
where $\{ , \}$ denote the Poisson bracket. ${\bf L}$ and ${\bf
A}$
 are first integrals of $H_{N}$, only $H_{1PN}+H_{2PN}+H_{S}$
 contributes to the right-hand sides of Eq.($\ref{e1.7}$) and
 Eq.($\ref{e1.8}$), in which $H_{1PN}+H_{2PN}$ determines the
 precession of periastron, in 1PN it is given,
\begin{equation}
\label{wex1} \dot{\omega}^{GR}=\frac {6\pi M}{P_b a(1-e^{2})} \ ,
\end{equation}
where $P_b$ is the orbital period. To study the spin-orbit
interaction, it is sufficient to consider $H_{S}$. Thus replacing
$H$ of Eq.($\ref{e1.7}$) and
 Eq.($\ref{e1.8}$) by $H_S$ obtains (Damour $\&$ Sch\"afer~1988),
\begin{equation}
\label{e1.13} (\frac{d{\bf L}}{dt})_{S}=\{{\bf L},H_{S}\}=\dot{
\Omega}_{S}^{\ast}\hat{{\bf S}}\times {\bf L}  \ ,
\end{equation}
\begin{equation}
\label{e1.14} (\frac{d{\bf A}}{dt})_{S}=\{{\bf A},H_{S}\}= \dot{
\Omega}_{S}^{\ast}[\hat{{\bf S}}-3(\hat{{\bf L}}\cdot \hat{{\bf
S}})\hat{{\bf L}}]\times {\bf A}  \ .
\end{equation}
where
\begin{equation}
\label{wex2}\dot{\Omega}_{S}^{\ast}=\frac
{S(4+3m_{2}/m_{1})}{2r^{3}} \ .
\end{equation}
By Damour $\&$ Sch\"afer~(1988), ${\bf S}$ represents a linear
combination of ${\bf S}_1$ and ${\bf S}_2$. For simplicity of
discussion, and for consistence with Wex $\&$ Kopeikin's
application of $\dot{\bf \Omega}_{S}$ (Wex $\&$ Kopeikin 1999), we
assume ${\bf S}={\bf S}_1$ (the other spin angular momentum is
ignored) until Sec~4 where the general binary pulsar is discussed.

The solution of Eq.($\ref{e1.13}$) and Eq.($\ref{e1.14}$) gives
the S-L coupling induced orbital precession velocity (Damour $\&$
Sch\"afer~1988)
\begin{equation}
\label{we3} \dot{\bf \Omega}_{S}=\dot{
\Omega}_{S}^{\ast}[\hat{{\bf S}}-3(\hat{{\bf L}}\cdot \hat{{\bf
S}})\hat{{\bf L}}] \,.
\end{equation}
By Eq.($\ref{e1.13}$) and Eq.($\ref{we3}$), the first derivative
of $\hat{\bf L}$ can be obtained,
\begin{equation}
\label{dotL} \frac{d\hat{\bf L}}{dt}=\dot{\Omega}^{\ast}_S\hat{\bf
S}\times \hat{\bf L} \ ,
\end{equation}
and by  Eq.($\ref{e1x1}$) the first derivative of $\hat{\bf S}$
(recall ${\bf S}={\bf S}_1$) can be written
\begin{equation}
\label{dotS} \frac{d\hat{\bf S}}{dt}=\dot{\Omega}^{\ast}_1\hat{\bf
L}\times \hat{\bf S}=\Omega^{\ast}_S\frac{L}{S}\hat{\bf L}\times
\hat{\bf S} \ ,
\end{equation}
where $\dot{\Omega}^{\ast}_1$ is the first term at the right hand
side of Eq.($\ref{e1x1}$). By Eq.($\ref{dotL}$) and
Eq.($\ref{dotS}$)  $\hat{\bf L}$ precesses slowly around $\hat{\bf
S}$, 1.5PN, as shown by Eq.($\ref{dotL}$); whereas $\hat{\bf S}$
precesses rapidly around $\hat{\bf L}$, 1PN, as shown in
Eq.($\ref{dotS}$). Therefore the BO equation predicts such a
scenario that the two vectors, $\hat{\bf L}$ and $\hat{\bf S}$
precess around each with very different precession velocities
(typically one is larger than the other by 3 to 4 orders of
magnitude for a general binary pulsar system).

\subsection{Orbital precession velocity in the calculation of
gravitational waves} In the study of the modulation of precession
of orbital plane to gravitational waves, the orbital precession
velocity is obtained in a different manner and the result is very
different from that given by Eq.($\ref{we3}$).

Since the gravitational wave corresponding to 2.5PN, which is
negligible compared with S-L coupling effect that corresponds to
1PN and 1.5PN, the total angular momentum can be treated as
conserved, $\dot{{\bf J}} = 0$. Then the following equation can be
obtained (BO),
\begin{equation}
\label{e1} \dot{\bf \Omega}_{0}\times {\bf L}=-\dot{\bf
\Omega}_{1}\times {\bf S}_{1}-\dot{\bf \Omega}_{2}\times{\bf
S}_{2}
 \ \,.
\end{equation}
Notice that as defined by BO and AK, ${\bf L}=\mu
M^{1/2}r^{1/2}\hat{\bf L}$. In the one-spin case the right hand
side of Eq.($\ref{e1}$) can be given (Kidder~1995),
\begin{equation}
\label{spin1} \dot{\bf S}=\frac {1}{2r^{3}}
(4+\frac{3m_{2}}{m_{1}}) ({\bf L}\times{\bf S}) \,,
\end{equation}
and considering that ${\bf J}={\bf L}+{\bf S}$, Eq.($\ref{spin1}$)
can be written,
\begin{equation}
\label{spin2} \dot{\bf S}=\frac {1}{2r^{3}}
(4+\frac{3m_{2}}{m_{1}}) ({\bf J}\times{\bf S}) \ \,.
\end{equation}
From Eq.($\ref{e1}$), $\dot{\bf L}$ can be given,
\begin{equation}
\label{spin3} \dot{\bf L}=\frac {1}{2r^{3}}
(4+\frac{3m_{2}}{m_{1}}) ({\bf J}\times{\bf L}) \ \,.
\end{equation}
By Eq.($\ref{spin2}$) and Eq.($\ref{spin3}$), ${\bf L}$ and ${\bf
S}$ precess about the fixed vector ${\bf J}$ at the same rate with
a precession frequency approximately (AK)
\begin{equation}
\label{e1aa4}\dot{\bf\Omega}_{0}=\frac {{\bf J}}{2r^{3}}
(4+\frac{3m_{2}}{m_{1}}) \ .
\end{equation}
Eq.($\ref{e1aa4}$) indicates that in the 1~PN approximation,
$\hat{\bf L}$ and $\hat{\bf S}$ can precess around ${\bf J}$
rapidly (1PN) with exactly the same velocity. Notice that the
misalignment angles between $\hat{\bf L}$ and $\hat{\bf S}$
($\lambda_{LS}$), $\hat{\bf L}$ and $\hat{\bf J}$ ($\lambda_{LJ}$)
are very different,  due to $S/L\ll 1$ , $\lambda_{LJ}$  is much
smaller than $\lambda_{LS}$.

Thus, AK's equations, Eq.($\ref{spin3}$) and Eq.($\ref{spin2}$),
correspond to a very different scenario of motion of ${\bf S}$,
${\bf L}$ and ${\bf J}$ from that given by BO equation shown in
Eq.($\ref{dotL}$) and Eq.($\ref{dotS}$).


\section{Physical differences between BO and AK}
This section compares two different scenarios corresponding to BO
and AK's orbital precession velocity, and points out that BO's
orbital precession velocity is  actually inconsistent with the
definition of the  total angular momentum of a binary system.

Section 2 indicates that BO and AK derived the orbital precession
velocity in different ways, therefore two different orbital
precession velocity vectors are  obtained, as shown in
Eq.($\ref{we3}$) and Eq.($\ref{e1aa4}$), respectively, which in
turn correspond to different scenarios of motion of the three
vectors. This section analyzes that the discrepancy between BO and
AK is not just a discrepancy corresponding to different coordinate
systems. Actually there is significant physical differences
between BO and AK. The total angular momentum of a binary system
is defined as BO
\begin{equation}
\label{triangle} {\bf J}={\bf L}+{\bf S} \ ,
\end{equation}
Eq.($\ref{triangle}$) means that ${\bf J}$, ${\bf L}$ and ${\bf
S}$ form a triangle, and therefore, it guarantees that the three
vectors must be in one plane at any moment. For a general radio
binary pulsar system, the total angular momentum of this system is
conserved in 1PN. Therefore we have
\begin{equation}
\label{dotj} \dot{\bf J}=0 \ .
\end{equation}
Eq.($\ref{dotj}$) means that ${\bf J}$ is a constant during the
motion of a binary system.  Eq.($\ref{dotj}$) and
Eq.($\ref{triangle}$) together provide a scenario that the
triangle formed by ${\bf L}$, ${\bf S}$ and ${\bf J}$ determines a
plane, and the plane rotates around a fixed axis, ${\bf J}$, with
velocity $\dot{\Omega}_0$.  This scenario is shown in Fig~1. which
can also be represented as
\begin{equation}
\label{dotjkid} \dot{\bf J}=\dot{\Omega}_0\hat{\bf J}\times{\bf
L}+\dot{\Omega}_0\hat{\bf J}\times{\bf S}=0 \,,
\end{equation}
Smarr $\&$ Blandford~(1976) mentioned the scenario that ${\bf L}$
and ${\bf S}$ must be at opposite side of ${\bf J}$ at any
instant. Hamilton and Sarazin~(1982) also study the scenario and
state that ${\bf L}$ precesses rapidly around ${\bf J}$. Obviously
the orbital precession velocity given by Eq.($\ref{e1aa4}$) can
satisfy the two constraints, Eq.($\ref{triangle}$) and
Eq.($\ref{dotj}$) simultaneously.

Can the  BO's orbital precession velocity given by
Eq.($\ref{we3}$) satisfy the two constraints,
Eq.($\ref{triangle}$) and Eq.($\ref{dotj}$) simultaneously ? From
Eq.($\ref{we3}$), Eq.($\ref{dotL}$) and Eq.($\ref{dotS}$), the
first derivative of ${\bf J}$ can be written (BO)
\begin{equation}
\label{bocons} \dot{\bf J}=\dot{\bf \Omega}_S\times {\bf
L}+\dot{\bf \Omega}^{\ast}_1\times {\bf S}
=\dot{\Omega}^{\ast}_S\hat{\bf S}\times {\bf L}+\dot{
\Omega}^{\ast}_1\hat{\bf L}\times {\bf S} \equiv 0  \ ,
\end{equation}
and since Eq.($\ref{triangle}$) is defined in BO's equation, then
it seems that the BO equation can satisfy both
Eq.($\ref{triangle}$) and Eq.($\ref{dotj}$).

But  in BO's derivation of $\dot{\bf \Omega}_S$
(Eq.($\ref{e1.7}$)--Eq.($\ref{we3}$)), Eq.($\ref{triangle}$) is
never used. The corresponding $\dot{\bf \Omega}_S$ can make
$\dot{\bf J}=\dot{\bf L}+\dot{\bf S}\equiv 0$, as shown in
Eq.($\ref{bocons}$), however it cannot guarantee that ${\bf
J}={\bf L}+{\bf S}$ is satisfied. In other words, when ${\bf
J}\neq{\bf L}+{\bf S}$, Eq.($\ref{bocons}$) is still correct. This
can be easily tested by putting ${\bf L}^{\prime}={\bf
L}+\alpha{\bf S}$, or ${\bf S}^{\prime}={\bf S}+\beta{\bf L}$
($\alpha$ and $\beta$ are arbitrary constants) into
Eq.($\ref{bocons}$) to replace ${\bf L}$ and ${\bf S}$,
respectively, obviously in such cases, Eq.($\ref{bocons}$) is
still satisfied ($\dot{\bf J}\equiv 0$).

Contrarily in AK's derivation of $\dot{\bf
\Omega}_0$(Eq.($\ref{spin1}$)--Eq.($\ref{spin3}$)), the relation
Eq.($\ref{triangle}$) is used. And if we do the same replacement
of ${\bf L}^{\prime}={\bf L}+\alpha{\bf S}$, or ${\bf
S}^{\prime}={\bf S}+\beta{\bf L}$ in  Eq.($\ref{dotjkid}$),  then
Eq.($\ref{dotjkid}$) is violated ($\dot{\bf J}\neq 0$). This means
that for AK's $\dot{\bf \Omega}_0$, if Eq.($\ref{triangle}$) is
violated then Eq.($\ref{dotj}$) is violated also. Thus in AK's
expression, the conservation of the total angular momentum is
dependent on Eq.($\ref{triangle}$), whereas, in BO's expression,
the conservation of the total angular momentum is independent of
Eq.($\ref{triangle}$). If we rewrite  Eq.($\ref{triangle}$) as,
\begin{equation}
\label{triangle2} {\bf J}={\bf L}+{\bf S}+{\bf C} \ ,
\end{equation}
then in BO's expression, the conservation of the total angular
momentum can be satisfied in the case that ${\bf C}\neq 0 $ in
Eq.($\ref{triangle2}$); whereas, in AK's expression, the
conservation of the total angular momentum is satisfied only when
${\bf C}=0 $ in Eq.($\ref{triangle2}$). This means that the
discrepancy between BO and AK's orbital precession velocity is
physical. It is not just different expression in different
coordinate systems or relative to different directions.

Moreover Eq.($\ref{dotjkid}$) and Eq.($\ref{e1aa4}$) correspond to
the following orbital precession velocity,
\begin{equation}
\label{we3x} \dot{\bf
\Omega}_{0}=\dot{\Omega}_{S}^{\ast}(\hat{{\bf
S}}+\frac{L}{S}\hat{{\bf L}}) \,.
\end{equation}
Obviously  Eq.($\ref{we3x}$) is not consistent with BO's
Eq.($\ref{we3}$), which demands that the coefficient of the
component along $\hat{\bf L}$ be $\gamma=-3(\hat{\bf
L}\cdot\hat{\bf S})$, instead of $\gamma=\frac{L}{S}$ as given by
Eq.($\ref{we3x}$).

In other words, once Eq.($\ref{triangle}$) is satisfied,  BO's
orbital precession velocity of Eq.($\ref{we3}$) must be violated.
Therefore, BO's orbital precession velocity cannot be consistent
with BO's definition, ${\bf J}={\bf L}+{\bf S}$.

Actually Eq.($\ref{we3x}$) can be consistent with
Eq.($\ref{e1.13}$),  however it is contradictory to
Eq.($\ref{e1.14}$). The reason of introducing Eq.($\ref{e1.14}$)
is that without it, Eq.($\ref{e1.13}$) alone cannot determine a
unique solution.

Whereas, Eq.($\ref{dotjkid}$) and Eq.($\ref{e1aa4}$) can be
regarded as solving this problem by  using Eq.($\ref{e1.13}$) and
Eq.($\ref{triangle}$) instead of Eq.($\ref{e1.13}$) and
Eq.($\ref{e1.14}$) to obtain the orbital precession velocity.

As defined in Eq.($\ref{e1.5}$) and Eq.($\ref{e1.6}$), ${\bf L}$
and ${\bf A}$ are vectors that are determined by different
elements in celestial mechanics,  ${\bf L}(\Omega, i)$ and ${\bf
A}(\Omega, i, \omega, e)$ respectively. And these two vectors
satisfy different physical constraints, i.e.,  ${\bf L}$ satisfies
Eq.($\ref{triangle}$) and Eq.($\ref{dotj}$), whereas,  ${\bf A}$
doesn't satisfy these two constraints.

Therefore, it is conceivable that ${\bf L}$ and  ${\bf A}$ should
correspond to different precession velocities,  as given by
Eq.($\ref{e1.13}$) and Eq.($\ref{e1.14}$), respectively. However
since the discrepancy is only in the $\hat{\bf L}$ component,
which does not influence the satisfaction of  the conservation
equation, Eq.($\ref{bocons}$), thus the discrepancy seems
unimportant. And therefore, the precession velocity of ${\bf L}$
is  treated equivalently to that of ${\bf A}$'s, thus the
components in $\hat{\bf L}$ are both treated as
$\gamma=-3(\hat{\bf L}\cdot\hat{\bf S})$. Whereas, as given by
Eq.($\ref{we3x}$), $\hat{\bf L}$ component must be
$\gamma=\frac{L}{S}$ if the triangle constraint is to be
satisfied. Therefore, the violation of the triangle constraint is
inevitable under the assumption that ${\bf L}$ and  ${\bf A}$
precess at the same velocity.

\section{S-L coupling induced effects derived under different degree of freedom}
 As analyzed in Sect 2 and Sect 3, whether the
triangle constraint is satisfied or not results discrepancy in the
orbital precession velocity. This section further analyzes that it
is the discrepancy in degree of freedom used by BO and AK that
leads to the violation or satisfaction  of the triangle
constraint.

To discuss the S-L coupling induced effects on observational
parameters, one need to obtain the variation of the six orbital
elements under S-L coupling. The way of doing this  is from the
Hamiltonian (corresponding to S-L coupling) to equation of motion,
and then through perturbation methods in celestial mechanics to
obtain S-L coupling induced variation of the six orbital elements,
and finally transform effects  to observer's coordinate system. In
this section this process is performed in the case that the free
vectors  of a binary system (with two spins) is 3, in which the
triangle constraint is satisfied.

It is convenient to study the motion of a binary system in such a
coordinate system (J-coordinate system), in which the total
angular momentum, ${\bf J}$ is along the z-axes and the invariance
plane is in the x-y plane. The J-coordinate system has two
advantages.

(a) Once  a binary pulsar system is given, $\lambda_{LJ}$, the
misalignment angle between ${\bf J}$ and ${\bf L}$, can be
estimated, from which  $\dot{\Omega}$ and $\dot{\omega}$ can be
obtained easily in the J-coordinate system, which are intrinsic to
a binary pulsar system.

(b) Moreover,  the J-coordinate system is static relative to the
line of sight (after counting out the proper motion). Therefore
transforming  parameters obtained in the J-coordinates system to
observer's coordinate system, S-L coupling induced effects can be
obtained reliably.

From Eq.($\ref{e1.2hs}$), the S-L coupling induced $H_S$ contains
potential part only, therefore we have $H_S=U$, where
\begin{equation}
\label{vv}  U=U_1+U_2=\frac{1}{r^3}(2{\bf S}+\frac{3m_2}{2m_1}{\bf
S}_1+\frac{3m_1}{2m_2}{\bf S}_2)\cdot {\bf L} \ ,
\end{equation}
which can be written as
\begin{equation}
\label{v}  U=\frac{1}{r^3}(\sigma_1{\bf S}\cdot {\bf
L}+\sigma_2{\bf S}_2\cdot {\bf L})
 \ \,,
\end{equation}
where
\begin{equation}
\label{sigma}\sigma_1=2+\frac{3}{2}\frac{m_2}{m_1}
 \ \  , \ \  \sigma_2=2+\frac{3}{2}(\frac{m_1}{m_2}-\frac{m_2}{m_1}) \ \
 .
\end{equation}
From Eq.($\ref{v}$) we can have the Lagrange corresponds to S-L
coupling, $\Im=-U$. And the Lagrange equation is
\begin{equation}
\label{lagrange}\frac{d}{dt}(\frac{\partial \Im}{\partial
\dot{q}_{\kappa}})-\frac{\partial \Im}{\partial q_{\kappa}}=0 \ \
 , \ \ (\kappa=1,2,...,\beta)
\end{equation}
where $ q_{\kappa}$ is the generalized coordinate ($\beta$ is the
number of  degrees of freedom), given by $r_1^{(\alpha)}$,
$r_2^{(\alpha)}$, $S_1^{(\alpha)}$, $S_2^{(\alpha)}$ (which
represent the position of body 1 and body 2; the spin angular
momentum of body 1 and body 2 respectively, $\alpha=1,2,3$).

Since the times scale of spin down of a pulsar in a binary pulsar
system is much longer than that of the period of geodetic
precession, the magnitude of ${\bf S}_1$ and ${\bf S}_2$ can be
treated as constant (these two vectors can only vary in
direction). However since the misalignment angle between ${\bf
S}_1$ and ${\bf L}$ ($\lambda_{LS1}$), as well as ${\bf S}_2$ and
${\bf L}$ ($\lambda_{LS2}$), are also constants in BO's
gravitational two-body equation (Barker $\&$ O'Connell 1975),
we have $\partial({\bf S}_1\cdot{\bf L})/\partial
S_1^{(\alpha)}=0$ and $\partial({\bf S}_2\cdot{\bf L})/\partial
S_2^{(\alpha)}=0$. Thus we have  ${\partial \Im}/{\partial
q_{\kappa}}=0$ for $q_{\kappa}=S_1^{(\alpha)}, S_2^{(\alpha)}$.
And since $\dot{\bf S}_1$ and $\dot{\bf S}_2$ are not appeared in
the Lagrange $\Im$, we have $d({\partial \Im}/{\partial
\dot{q}_{\kappa}})/dt=0$ for $q_{\kappa}=S_1^{(\alpha)},
S_2^{(\alpha)}$.

Therefore, we only need to calculate $d({\partial \Im}/{\partial
\dot{q}_{\kappa}})/dt$ and ${\partial \Im}/{\partial q_{\kappa}}$
of  Eq.($\ref{lagrange}$) in the case that
$q_{\kappa}=r_1^{(\alpha)}, r_2^{(\alpha)}$ .
 The first term at the left hand
side  of  Eq.($\ref{lagrange}$) corresponds to generalized force,
which can be written $d({\partial \Im}/{\partial
\dot{q}_{\kappa}})/dt=F=\mu a_{so}$; and the second term is
${\partial \Im}/{\partial q_{\kappa}}=-\nabla U$ (where $\nabla$
represents gradient). Thus Eq.($\ref{lagrange}$) can be rewritten,
\begin{equation}
\label{asov}{\bf a}_{so}=-\frac{1}{\mu}\nabla
U=-\frac{1}{\mu}[{\sigma_1}\nabla \frac{({\bf S}\cdot{\bf
L})}{r^3}+{\sigma_2}\nabla \frac{({\bf S}_2\cdot{\bf L})}{r^3}] \
\ ,
\end{equation}
The triangle constraint  given by Eq.($\ref{triangle}$) indicates
that ${\bf S}$ and ${\bf L}$ are dependent. Therefore, we cannot
treat the free variables of Eq.($\ref{asov}$) as $r_1^{(\alpha)}$,
$r_2^{(\alpha)}$, $S_1^{(\alpha)}$ and $S_2^{(\alpha)}$.

Classical mechanics tells us that  constraints actually reduce the
number of degrees of freedom of a dynamic system. The geometric
constraint  ${\bf J}={\bf L}+{\bf S}$ can be imposed in
Eq.($\ref{asov}$) through the replacement,  ${\bf S}={\bf J}-{\bf
L}$. Thus the free variables in Eq.($\ref{asov}$) is either
$r_1^{(\alpha)}$, $r_2^{(\alpha)}$, and $S_1^{(\alpha)}$; or
$r_1^{(\alpha)}$, $r_2^{(\alpha)}$, and $S_2^{(\alpha)}$
(depending on different definition of $\sigma_1$ and $\sigma_2$ in
Eq.($\ref{sigma}$)). Therefore, performing the replacement ${\bf
S}={\bf J}-{\bf L}$ in Eq.($\ref{asov}$) actually means that the
triangle constraint is imposed on the motion of binary system, and
the degrees of freedom are reduced from  12 ($4\times 3$) to 9
($3\times 3$).

Considering that ${\bf S}_1$ and ${\bf S}_2$ can vary in direction
only, $\alpha$ of  $S_1^{(\alpha)}$, $S_2^{(\alpha)}$ is given
$\alpha=1,2$. Thus the degree of freedom of a binary pulsar system
without the triangle constraint is 10, and with the constraint is
8.

This is analogy to the calculation of equation of  motion of a
simple clock pendulum. The motion of a small mass at the bottom of
a clock pendulum can be described in $x-y$ plane. However if we
treat the degrees of freedom of this small mass as 2, then this
small mass can move every in the 2-dimension space, and the length
of the pendulum is not a constant. In other words, once the length
of the pendulum is a constant, the degree of freedom is 1 instead
of 2. Correspondingly if the degree of freedom of a binary system
is 10, then the satisfaction of the triangle constraint cannot be
guaranteed (or ${\bf J}$=const vector cannot be guaranteed).
Contrarily, if the triangle constraint is satisfied, the degree of
freedom 8 instead of 10. By the replacement, ${\bf S}={\bf J}-{\bf
L}$, Eq.($\ref{asov}$) can be re-written
\begin{equation}
\label{LSJ}\nabla \frac{({\bf S}\cdot{\bf
L})}{r^3}=\nabla\frac{[({\bf J}-{\bf L})\cdot{\bf
L}]}{r^3}=\nabla\frac{({\bf J}\cdot{\bf
L})}{r^3}-\nabla\frac{({\bf L}\cdot{\bf L})}{r^3} \ \ ,
\end{equation}
where
\begin{equation}
\label{LSJ1} \nabla\frac{({\bf J}\cdot{\bf L})}{r^3}=
 -3\frac{({\bf L}\cdot{\bf J}){\bf r}}{r^5}-3\frac{({\bf J}\times{\bf
r})({\bf V}\cdot{\bf r})}{r^5}\mu+2\frac{({\bf J}\times{\bf
V})}{r^3}\mu \ \,,
\end{equation}
\begin{equation}
\label{LSJ1a} \nabla\frac{({\bf L}\cdot{\bf L})}{r^3}=
 -3\frac{({\bf L}\cdot{\bf L}){\bf r}}{r^5}-2\frac{({\bf L}\times{\bf V})}{r^3}\mu
\ \,,
\end{equation}
where  ${\bf V}$ is the velocity of the reduced mass. Put
Eq.($\ref{LSJ1}$) and Eq.($\ref{LSJ1a}$) into Eq.($\ref{LSJ}$) and
finally into  Eq.($\ref{asov}$), we have,
$${\bf a}_{so}=\frac{3}{r^3}[\sigma_1 ({\bf J}-{\bf L})\cdot(\hat{\bf n}\times{\bf
V})\hat{\bf n}+\sigma_ 2{\bf S}_2\cdot(\hat{\bf n}\times{\bf
V})\hat{\bf n}] $$
$$+\frac{2}{r^3}[2\sigma_1 ({\bf V}\times{\bf
J})-\sigma_ 1({\bf V}\times({\bf J}-{\bf L}))+\sigma_ 2({\bf
V}\times{\bf S}_2)]
$$
\begin{equation}
\label{asoo0} +\frac{3({\bf V}\cdot\hat{\bf
n})}{r^3}[\sigma_1({\bf J}\times\hat{\bf n})+\sigma_2({\bf
S}_2\times\hat{\bf n})] \ \,,
\end{equation}
where $\hat{\bf n}$ is the unit vector of ${\bf r}$. Replacing
${\bf J}-{\bf L}$ by ${\bf S}$, Eq.($\ref{asoo0}$) can be written,
$${\bf a}_{so}=\frac{3}{r^3}[\sigma_1 {\bf S}\cdot(\hat{\bf n}\times{\bf
V})\hat{\bf n}+\sigma_ 2{\bf S}_2\cdot(\hat{\bf n}\times{\bf
V})\hat{\bf n}] $$
$$+\frac{2}{r^3}[2\sigma_1 ({\bf V}\times{\bf
J})-\sigma_ 1({\bf V}\times{\bf S})+\sigma_ 2({\bf V}\times{\bf
S}_2)]
$$
\begin{equation}
\label{asoo} +\frac{3({\bf V}\cdot\hat{\bf n})}{r^3}[\sigma_1({\bf
J}\times\hat{\bf n})+\sigma_2({\bf S}_2\times\hat{\bf n})] \ \,.
\end{equation}
If one calculates ${\bf a}_{so}$ directly by Eq.($\ref{asov}$)
without imposing the triangle constraint, then the corresponding
result can be  given by replacing ${\bf J}$ of Eq.($\ref{asoo}$)
by ${\bf S}$,
$${\bf a}_{so}^{\prime}=\frac{3}{r^3}[\sigma_1 {\bf S}\cdot(\hat{\bf n}\times{\bf
V})\hat{\bf n}+\sigma_ 2{\bf S}_2\cdot(\hat{\bf n}\times{\bf
V})\hat{\bf n}] $$
$$+\frac{2}{r^3}[\sigma_ 1({\bf V}\times{\bf S})+\sigma_ 2({\bf V}\times{\bf
S}_2)]
$$
\begin{equation}
\label{asoox} +\frac{3({\bf V}\cdot\hat{\bf
n})}{r^3}[\sigma_1({\bf S}\times\hat{\bf n})+\sigma_2({\bf
S}_2\times\hat{\bf n})] \ \,.
\end{equation}
Notice that Eq.($\ref{asoox}$) is equivalent to the sum of Eq.(52)
and Eq.(53) given by the BO equation. The difference between
Eq.($\ref{asoo}$) and Eq.($\ref{asoox}$) indicates that whether
the triangle constraint is satisfied or not  can lead to
significant differences in ${\bf a}_{so}$, which in turn results
in significant differences on the predictions of observational
effects as discussed in the next section.

Having ${\bf a}_{so}$, we can use the standard method in celestial
mechanics to calculate,
\begin{equation}
\label{stw} \widetilde{S}={\bf a}_{so}\cdot\hat{\bf n}, \ \
\widetilde{T}={\bf a}_{so}\cdot\hat{\bf t}, \ \ \widetilde{W}={\bf
a}_{so}\cdot\hat{\bf L} ,
\end{equation}
from which one can calculate the derivative of the six orbit
elements and then transform to the observer's coordinate system to
compare with observation. The unit vector, $\hat{\bf n}$, in
Eq.($\ref{asoo}$) to Eq.($\ref{stw}$)   is given by,
\begin{equation}
\label{n}\hat{\bf n}={\bf P}\cos f+{\bf Q}\sin f \ \,,
\end{equation}
and $\hat{\bf t}$ is the unit vector that is perpendicular to $
\hat{\bf n}$,
\begin{equation}
\label{t}\hat{\bf t}=-{\bf P}\sin f+{\bf Q}\cos f \ \,,
\end{equation}
and  ${\bf V}={\bf p}/\mu$, is given by
\begin{equation}
\label{V}{\bf V}=-\frac{h}{p}{\bf P}\cos f+\frac{h}{p}{\bf
Q}(e+\cos f) \ \,,
\end{equation}
where $f$ is the true anomaly, $p$ is the semilatus rectum,
$p=a(1-e^2)$, and $h$ is the integral of area, $h=r^2\dot{f}$.
${\bf P}$ is given by three components,
$$ P_x= \cos\Omega\cos\omega-\sin\Omega\sin\omega\cos\lambda_{LJ}  \ \ ,$$
$$
P_y=\sin\Omega\cos\omega+\cos\Omega\sin\omega\cos\lambda_{LJ} \ \
,
$$
\begin{equation}
\label{P} P_z= \sin\omega\sin\lambda_{LJ} \ \ ,
\end{equation}
and ${\bf Q}$ is given by three components,
$$ Q_x= -\cos\Omega\sin\omega-\sin\Omega\cos\omega\cos\lambda_{LJ}  \ \ ,$$
$$
Q_y=-\sin\Omega\sin\omega+\cos\Omega\cos\omega\cos\lambda_{LJ} \ \
,
$$
\begin{equation}
\label{Q} Q_z= \cos\omega\sin \lambda_{LJ} \ \ ,
\end{equation}
The unit vector of $\hat{\bf L}$ and $\hat{\bf S}_\kappa$
($\kappa=1,2$) are given,
\begin{equation}
\label{L} \hat{\bf L}=(\sin\lambda_{LJ}\cos\eta_L, \
\sin\lambda_{LJ}\sin\eta_L, \ \cos\lambda_{LJ})^T \ \ ,
\end{equation}
\begin{equation}
\label{S2} \hat{\bf
S}_{\kappa}=(\sin\lambda_{JS\kappa}\cos\eta_{S\kappa}, \
\sin\lambda_{JS\kappa}\sin\eta_{S\kappa}, \
\cos\lambda_{JS\kappa})^T \ \ .
\end{equation}
In the perturbation equation, the acceleration of
Eq.($\ref{asoo}$), ${\bf a}_{so}$, is expressed along  $\hat{\bf
n}$, $\hat{\bf t}$ and $\hat{\bf L}$ respectively. We can use
${\bf a}_1$, ${\bf a}_2$ and ${\bf a}_3$ to represent terms
corresponding to the three terms containing brackets $[,]$  at the
right hand side of Eq.($\ref{asoo}$), respectively.
Projecting ${\bf a}_1$ onto $\hat{\bf L}$, we have
$$ W_1={\bf a}_1\cdot \hat{\bf L}=\frac{3\sigma_1}{r^3}
[S_x(n_yV_z-n_zV_y)
+S_y(n_zV_x-n_xV_z)+S_z(n_zV_y-n_yV_z)]$$
\begin{equation} \label{w2a}
(n_x\sin \lambda_{LJ}\cos\eta_L+n_y\sin
\lambda_{LJ}\sin\eta_L+n_z\cos\lambda_{LJ}) \ ,
\end{equation}
where $n_x$,  $n_y$, $n_z$ and $V_x$,  $V_y$, $V_z$ are components
of $\hat{\bf n}$ and  ${\bf V}$ along axes, $x$, $y$ and $z$,
respectively. Similarly, projecting ${\bf a}_2$ onto $\hat{\bf
L}$, we have,
$$ W_2={\bf a}_2\cdot \hat{\bf L}=\frac{\sigma_1}{r^3}
(V_yJ\sin\lambda_{LJ}\cos\eta_L-V_xJ\sin\lambda_{LJ}\sin\eta_L)$$
\begin{equation} \label{w2ax}
+[\frac{\sigma_1}{r^3}(V_xS_{y}-V_yS_{x})
+\frac{2\sigma_2}{r^3}(V_xS_{2y}-V_yS_{2x})]\cos \lambda_{LJ} \ .
\end{equation}
Finally  ${\bf a}_3$ can also be projected onto $\hat{\bf L}$,
\begin{equation} \label{w3a}
W_3={\bf a}_3\cdot \hat{\bf
L}=\frac{\cos\lambda_{LJ}}{r^3}[{\sigma_1}(V^r_yS_x-V^r_xS_y)
 +{\sigma_2}(V^r_yS_{2x}-V^r_xS_{2y})] \ ,
\end{equation}
where $V^r_x=3\dot{r}n_x$, $V^r_y=3\dot{r}n_y$ and
$\dot{r}=\frac{eh}{p}\sin f$. Therefore, the sum of W is
\begin{equation} \label{w3b}
 \widetilde{W}=W_1+W_2+W_3 \ .
\end{equation}
The effect around ${\bf J}$ can be obtained by perturbation
equations (Roy~1991, Yi~1993, Liu~1993) and Eq.($\ref{w3b}$)
\begin{equation} \label{Womega}
\frac{d\Omega}{dt}=\frac{\widetilde{W}r\sin
(\omega+f)}{na^2\sqrt{1-e^2}}\frac{1}{\sin\lambda_{LJ}} \ ,
\end{equation}
where $n$ is the angular velocity. Averaging over one orbital
period we have
$$
<\frac{d\Omega}{dt}>=\frac{3\cos\lambda_{LJ}
}{2a^3(1-e^2)^{3/2}\sin\lambda_{LJ}}(P_z\sin\omega+Q_x\cos\omega)
[(P_yQ_z-P_zQ_y)(S_x\sigma_1+S_{2x}\sigma_2)$$
\begin{equation} \label{Womegaav}
+(P_zQ_x-P_xQ_z)(S_y\sigma_1+S_{2y}\sigma_2)
+(P_xQ_y-P_yQ_x)(S_z\sigma_1+S_{2z}\sigma_2)] \ .
\end{equation}
Notice that the average value of  Eq.($\ref{Womegaav}$) depends on
$W_1$ only, the contribution of   $W_2$ and  $W_3$ to it is zero.
With $S/\sin\lambda_{LJ}\sim L$,  we have ${d\Omega}/{dt}\sim
{L}/{a^3}$, which corresponds to 1PN.

The  $d\omega/dt$ can be obtained  by calculation of
$\widetilde{S}={\bf a}_{so}\cdot \hat{\bf n}$ and
$\widetilde{T}={\bf a}_{so}\cdot \hat{\bf t}$. Since ${\bf
a}_{1}\cdot \hat{\bf n}$ and ${\bf a}_{1}\cdot \hat{\bf t}$ are
1.5PN. Therefore, it is sufficient to consider the projection of
${\bf a}_2$  and ${\bf a}_3$ onto $\hat{\bf n}$, $\hat{\bf t}$
respectively, thus we have,
\begin{equation} \label{a2S}  <({\bf a}_2\cdot \hat{\bf
n})\cos f>=\frac{7\sigma_1
J}{8(1-e^2)^{3/2}a^3}\frac{eh}{p}(P_xQ_y-P_yQ_x) \ \ ,
\end{equation}
\begin{equation} \label{as2V} <({\bf a}_2\cdot \hat{\bf
t})\sin f>=\frac{-\sigma_1
J}{2(1-e^2)^{3/2}a^3}\frac{eh}{p}(P_xQ_y-P_yQ_x) \ \ ,
\end{equation}
\begin{equation} \label{as3t} <({\bf a}_3\cdot \hat{\bf
t})\sin f>=\frac{-3\sigma_1
J}{2(1-e^2)^{3/2}a^3}\frac{eh}{p}(P_xQ_y-P_yQ_x) \ \ ,
\end{equation}
\begin{equation} \label{as3n} <({\bf a}_3\cdot \hat{\bf
n})\cos f>=0 \ \ .
\end{equation}
From Eq.($\ref{a2S}$) to Eq.($\ref{as3n}$), we have
$$
\frac{d\omega^{\prime}}{dt}=\frac{\sqrt{1-e^2}}{nae} \{[-{\bf
a}_2\cdot \hat{\bf n}]\cos f +(1+\frac{r}{p})[{\bf a}_2\cdot
\hat{\bf t}]\sin f \}$$
\begin{equation} \label{STomega0}
=\frac{7\sigma_1 J}{2(1-e^2)^{3/2}a^3}(P_yQ_x-P_xQ_y) \ \ .
\end{equation}
therefore, by the standard perturbation(Roy 1991, Yi 1993, Liu
1993), the advance of precession of periastron induced by S-L
coupling is given by
\begin{equation} \label{STomega0a}\frac{d\omega}{dt}
=\frac{d\omega^{\prime}}{dt}-\frac{d\Omega}{dt}\cos \lambda_{LJ} \
\ .
\end{equation}
By putting Eq.($\ref{a2S}$) and Eq.($\ref{as2V}$) into
Eq.($\ref{STomega0a}$), and averaging over one orbital period we
have,
\begin{equation} \label{STomega}
<\frac{d\omega}{dt}>=\frac{7\sigma_1
J}{2(1-e^2)^{3/2}a^3}(P_yQ_x-P_xQ_y) -\frac{d\Omega}{dt}\cos
\lambda_{LJ} \ \ .
\end{equation}
Using perturbation equations as in (Roy~1991, Yi~1993, Liu~1993),
and by Eq.($\ref{Womega}$) and Eq.($\ref{STomega}$), we have
$$
<\frac{d\varpi}{dt}>= \frac{7\sigma_1
J}{2(1-e^2)^{3/2}a^3}(P_yQ_x-P_xQ_y)
+2\frac{d\Omega}{dt}\sin^2\frac{\lambda_{LJ}}{2}
$$
\begin{equation} \label{SOomega}
=\frac{7\sigma_1 J}{2(1-e^2)^{3/2}a^3}(P_yQ_x-P_xQ_y) +O(c^{-3})
  \ \ .
\end{equation}
Eq.($\ref{Womegaav}$), Eq.($\ref{STomega}$) and
Eq.($\ref{SOomega}$) indicate that the magnitude of
${d\Omega}/{dt}$, and ${d\varpi}/{dt}$ are both ${L}/{a^3}$ (1PN),
whereas, ${d\omega}/{dt}$ can be 1.5PN (or zero in 1PN) by
Eq.($\ref{STomega}$).

${d\Omega}/{dt}$ (1PN) of Eq.($\ref{Womega}$) is equivalent to
$\dot{\Phi}_S$ (1PN) which is given by  Wex $\&$ Kopeikin~(1999).
This is because the averaged value of ${d\Omega}/{dt}$ depends
only on ${\bf a}_{1}$, the first term containing bracket [,] in
${\bf a}_{so}$, as shown in Eq.($\ref{asoo}$). And both this paper
(${\bf a}_{so}$ of Eq.($\ref{asoo}$)) and that of BO equation
(${\bf a}_{so}^{\prime}$ of Eq.($\ref{asoox}$)) give the same
${\bf a}_{1}$. Thus, different authors give the equivalent value
on the averaged ${d\Omega}/{dt}$.

Whereas, ${d\omega}/{dt}$  of Eq.($\ref{STomega}$) and
$\dot{\Psi}_S$ (1PN) given by Wex $\&$ Kopeikin~(1999) are very
different in magnitude. The difference is due to the fact that
${d\omega}/{dt}$ given by Eq.($\ref{STomega}$) of this paper is
obtained by the ${\bf a}_{so}$ of Eq.($\ref{asoo}$); whereas, the
corresponding ${d\omega}/{dt}$ of Wex and Kopeikin~(1999) is
obtained by the ${\bf a}_{so}^{\prime}$ of Eq.($\ref{asoox}$),
which is equivalent of replacing $J$ of Eq.($\ref{asoo}$) by $S$.

And in turn, the difference between ${\bf a}_{so}$ and ${\bf
a}_{so}^{\prime}$ is due to that ${\bf a}_{so}$ satisfies the
triangle constraint; whereas ${\bf a}_{so}^{\prime}$  doesn't.
Therefore, a small difference in the equation of motion causes
significant discrepancy in the variation of elements, such as
${d\omega}/{dt}$.

\section{Effects on $\dot{\omega}$, $\dot{x}$, $\dot{P}_b$}
As shown in Sect~4 the S-L coupling effect can be treated as a
perturbation to the Newtonian two-body problem, and by the
standard method in celestial mechanics the variation of six
orbital elements can be obtained. This section calculates  what
$\dot{\omega}$, $\dot{x}$, $\dot{P}_b$ are for an observer when
the variation of the six orbital elements is given. The results of
this section is independent of the degree of freedom used in a
binary system.

The observational effect is studies in such a coordinate system
that the vector ${\bf K}_0$ (corresponding to line of sight) is
along the $z^{\prime}$-axis; the $x^{\prime}$ axis is along the
intersection of the plane of the sky and the invariance plane; and
the $y^{\prime}$-axis is perpendicular to $x^{\prime}-z^{\prime}$
plane, as shown in Fig~2a. Obviously this coordinate system is at
rest to "an observer" at the SSB. The relationship of dynamical
longitude of the ascending node, $\Omega$, dynamical longitude of
the periastron, $\omega$, and the orbital inclination, $i$,  can
be given
(Smarr $\&$ Blandford 1976, Wex $\&$ Kopeikin 1999)
\begin{equation}
\label{ef6} \cos i=\cos I\cos\lambda_{LJ}-\sin\lambda_{LJ}\sin
I\cos\Omega \ \,,
\end{equation}
and
$$ \sin i\sin\omega^{\rm obs}=(\cos
I\sin\lambda_{LJ}+\cos\lambda_{LJ}\sin I\cos\Omega)\sin\omega $$
\begin{equation}
\label{ef7}
 +\sin I \sin\Omega\cos\omega \ \,,
\end{equation}
$$ \sin i\cos\omega^{\rm obs}=(\cos I\sin\lambda_{LJ}+\cos\lambda_{LJ}\sin
I\cos\Omega)\cos\omega $$
\begin{equation}
\label{ef8} -\sin I \sin\Omega\sin\omega \ \,,
\end{equation}
where $I$ is the misalignment angle between ${\bf J}$ and the line
of sight. The semi-major axis of the pulsar is  defined as
\begin{equation}
\label{dotxdef} x\equiv \frac{a_p\sin i}{c}  \ \,.
\end{equation}
where $a_p$ is the semi-major axis of the pulsar.  By
Eq.($\ref{ef6}$), we have,
\begin{equation}
\label{dotx1}\dot{x}_1=\frac{a_p\cos
i}{c}\frac{di}{dt}=-x\dot{\Omega}\sin \lambda_{LJ}\sin\Omega\cot i
\ \,.
\end{equation}
The semi-major axis of the orbit is  $a=\frac{M}{m_2}a_p$, and
since the L-S coupling induced  $\dot{a}$ is a function of
$\Omega$ and $\omega$, as shown in the appendix,  we have
\begin{equation}
\label{dotx2}\dot{x}_2=\frac{\dot{a}_p\sin
i}{c}=\frac{\dot{a}}{a}x
 \ \,.
\end{equation}
 Therefore, the L-S coupling induced
$\dot{x}$ is given by
\begin{equation}
\label{e3c}\dot{x}=\dot{x}_1+\dot{x}_2=-x\dot{\Omega}\sin
\lambda_{LJ}\sin\Omega\cot i+\frac{\dot{a}}{a}x
 \ \,.
\end{equation}
By Eq.($\ref{e3c}$) we have
\begin{equation}
\label{e3cc}\ddot{x}=\dot{x}_1\left(\frac{\ddot{\Omega}}{\dot{\Omega}}+
\dot{\Omega}\cot\Omega+
\frac{\dot{\lambda}_{LJ}\cos{\lambda}_{LJ}}{\sin{\lambda}_{LJ}}\right)+
x\frac{\ddot{a}a-\dot{a}^2}{a^2}+\dot{x}_2\frac{\dot{a}}{a}
 \ .
\end{equation}
Notice that $\dot{\Omega}$ and $\ddot{\Omega}$  can be obtained by
Eq.($\ref{Womega}$). Considering $\lambda_{LJ}\ll 1$ and by
Eq.($\ref{ef7}$), Eq.($\ref{ef8}$), the observational advance of
precession of periastron is given
(Smarr $\&$ Blandford 1976, Wex $\&$ Kopeikin 1999),
\begin{equation}
\label{ef9a} {\omega}^{\rm obs}={\omega}+{\Omega}-\lambda_{LJ}\cot
i\sin\Omega \ \,.
\end{equation}
Therefore, we have
\begin{equation}
\label{ef9} \dot{\omega}^{\rm
obs}=\dot{\omega}+\dot{\Omega}-\dot{\lambda}_{LJ} \cot i\sin\Omega
\ \,.
\end{equation}
If $\dot{\omega}$ and $\dot{\Omega}$ are caused only by the S-L
coupling effect ($H=H_S$), then $\dot{\omega}$ is given by
Eq.($\ref{STomega}$). If we consider all terms of the Hamiltonian,
as given by Eq.($\ref{e1.2}$), then $\dot{\omega}$ should include
$\dot{\omega}^{GR}$, the advance of precession of periastron
predicted by general relativity, which caused by $H_{1PN}$ and
$H_{2PN}$. In such case $\dot{\omega}$ in Eq.($\ref{ef9}$) is
replaced by $\dot{\omega}^{GR}+\dot{\omega}$.
 Thus Eq.($\ref{ef9}$) can be written as
\begin{equation}
\label{ef10} \dot{\omega}^{\rm
obs}=\dot{\omega}^{GR}+\dot{\omega}^{S}
 \ \,.
\end{equation}
where
\begin{equation}
\label{ef10ad}
\dot{\omega}^{S}=\dot{\omega}+\dot{\Omega}-\dot{\lambda}_{LJ} \cot
i\sin\Omega\ \,.
\end{equation}
Notice that $\dot{\omega}^{S}$ is a function of time due to
$\dot{\Omega}$, $\dot{\omega}$ and $\lambda_{LJ}$ are function of
time, as shown in Eq.($\ref{Womegaav}$), Eq.($\ref{STomega}$)  and
Eq.($\ref{Womegaavi}$) respectively. Whereas, $\dot{\omega}^{GR}$
is a constant as shown in Eq.($\ref{wex1}$).

For a binary pulsar system with negligibly small eccentricity, the
effect of the  variation in the advance of periastron, $\omega$,
is absorbed by the redefinition of the orbital frequency. As
discussed by Kopeikin~(1996), $\omega^{\rm obs}+A_e(u)$ is given
\begin{equation}
\label{ef11} \omega^{\rm
obs}+A_e(u)=\omega_0+\frac{2\pi}{P_b}(t-t_0) \ \,,
\end{equation}
where $A_e(u)$ is the true anomaly, related to the eccentric
anomaly, $u$, by the  transcendental equation, $\omega_0$ is the
orbital phase at the initial epoch $t_0$.

At the time interval,  $\delta t=(t-t_0)$, there is a
corresponding $\delta\omega^{\rm obs}$ which causes a
corresponding $\delta P_b$ at the right hand side of
Eq.($\ref{ef11}$), therefore, ${P}_b$ is a function of time. Thus
we have
\begin{equation}
\label{ef11a} \delta\omega^{\rm obs}+A_e(u)=\frac{2\pi}{P_b}\delta
t \ \,.
\end{equation}
Write $1/P_b$ in Taylor series, we have
\begin{equation}
\label{taypb}\frac{1}{P_b}=\frac{1}{P_b(t_0)}-\frac{\dot{P}_b(t_0)\delta
t}{P_b^2(t_0)}+...
\end{equation}
Considering  $A_e(u)={2\pi}\delta t/{P_b(t_0)} $ and by
Eq.($\ref{ef11a}$), Eq.($\ref{taypb}$) obtains,
\begin{equation}
\label{ef11b} \delta\dot{\omega}^{\rm
obs}=-\frac{2\pi\dot{P}_b}{P^{2}_b}\delta t \ \,.
\end{equation}
Since $\dot{\omega}^{S}$ is a function of time, whereas
$\dot{\omega}^{GR}=$const, then we have $\ddot{\omega}^{\rm
obs}=\ddot{\omega}^{S}$ by Eq.($\ref{ef10}$). Assume
$F=\dot{\omega}^{S}$, and write it in Taylor series as:
$F=F_0+\dot{F}\delta t+\frac{1}{2}\ddot{F}\delta t^2$, obtains
$\delta F= \delta\dot{\omega}^{S}\approx\dot{F}\delta
t=\ddot{\omega}^{S}\delta t$. Therefore, $\delta\dot{\omega}^{\rm
obs}$ of Eq.($\ref{ef11b}$) becomes
$\delta\dot{\omega}^{S}=\ddot{\omega}^{S}\delta t$, from which
Eq.($\ref{ef11b}$)  can be written as
\begin{equation}
\label{ez2} \dot{P}_{b}=-\frac{\ddot{\omega}^{S}P_{b}^{2}}{2\pi} \
\,.
\end{equation}
By Eq.($\ref{ez2}$), the derivatives of $P_b$ can be obtained,
\begin{equation}
\label{ez2a}
 \ddot{P}_{b}=
\frac{2\dot{P}_b^2}{P_b}- \frac{P_b^2}{2\pi
}\frac{d^3{\omega}^{S}}{dt^3}\approx-\frac{P_b^2}{2\pi
}\frac{d^3{\omega}^{S}}{dt^3}
 \,,
\end{equation}
\begin{equation}
\label{ez2b}
 \frac{d^3{P}_b}{dt^3}\approx - \frac{P_b^2}{2\pi}\frac{d^4{\omega}^{S}}{dt^4}
\,.
\end{equation}

\section{Comparison of three different S-L coupling induced
$\dot{\omega}^{obs}$ and $\dot{P}_b$}
\subsection{Discrepancy between Wex $\&$ Kopeikin and this paper}
Sect~4 calculates the S-L coupling induced change of orbital
elements of a binary system directly in the J-coordinate system
(in which $z$ axis is along $\hat{\bf J}$ and $x-y$ plane is the
invariance plane). And Sect~5 transform the effects in
J-coordinate system to the observer's coordinate system, and
obtains $\dot{x}$, $\dot{\omega}^{obs}$ and $\dot{P}_b$. In which
${\omega}^{obs}$ is  equivalent to the definition of Wex $\&$
Kopeikin (1999) as shown in Fig~2a.

By Eq($\ref{STomega}$) the $\dot{\omega}$ can be 1.5 PN (or
$\dot{\omega}=0$ in 1PN), thus by Eq($\ref{ef10ad}$) the S-L
coupling induced precession of periastron, $\dot{\omega}^{S}$,
becomes
\begin{equation}
\label{wkdsme1}
\dot{\omega}^{S}\approx\dot{\Omega}-\dot{\lambda}_{LJ} \cot
i\sin\Omega  \ \,.
\end{equation}
Eq($\ref{wkdsme1}$) is the result corresponding to 8 degrees of
freedom, and $\dot{\omega}^{S}$ is 1PN. On the other hand,  Wex
$\&$ Kopeikin (1999), rewrite BO's orbital precession velocity,
Eq($\ref{we3}$) in the J-coordinate system, and then obtain
$\dot{x}$, $\dot{\omega}^{obs}$ in observer's coordinate system.

However the difference is that for Wex $\&$ Kopeikin (1999), all
the results in the J-coordinate system is calculated in to 10
degrees of freedom. In such case $J$ in Eq($\ref{STomega}$) is
replaced by $S$, thus the first term at the right hand side of
Eq($\ref{STomega}$) is 0.5PN smaller than that of the second term.
Therefore $\dot{\omega}$ can be represented by the second term at
the right hand side of Eq($\ref{STomega}$), which is  1PN. By
Eq($\ref{ef10ad}$), and considering
$\dot{\Omega}-\dot{\Omega}\cos\lambda_{LJ}\sim 1.5PN$
($\lambda_{LJ}\ll 1$), we have
\begin{equation}
\label{wkdsme2} \dot{\omega}^{S}\approx-\dot{\lambda}_{LJ} \cot
i\sin\Omega  \ \,.
\end{equation}
Although ${\lambda}_{LJ}\sim S/L\ll 1$, $\dot{\lambda}_{LJ}$ is
significant which is 1PN  ($\dot{\lambda}_{LJ}\sim \dot{\Omega}$),
as shown in Eq($\ref{didt}$) and Eq($\ref{Womegaavi}$).
Consequently  $\dot{\omega}^{S}$ is also 1PN, which leads to
significant $\ddot{\omega}^{S}$ (1PN), and therefore, significant
derivative of ${P}_b$ by Eq($\ref{ez2}$)-Eq($\ref{ez2b}$). In
other words Wex $\&$ Kopeikin's ${\omega}$ of Eq(59) (Wex $\&$
Kopeikin 1999) actually corresponding significant variabilities,
such as $\dot{\omega}^{S}$ and  $\dot{P}_b$, these results seem
ignored.

Therefore, this paper and Wex Kopeikin (1999), which corresponding
to  Eq($\ref{wkdsme1}$) and Eq($\ref{wkdsme2}$) respectively, both
predict to significant $\dot{\omega}^{S}$ and $\dot{P}_b$. However
there is an obvious discrepancy between them, that is in
Eq($\ref{wkdsme2}$), $\dot{\omega}^{S}\rightarrow 0$ when
$i\rightarrow \pi/2$. Whereas, Eq($\ref{wkdsme1}$) doesn't has
such a relation. Therefore, the validity of Wex $\&$ Kopeikin
(1999) and this paper can be tested by binary pulsar systems with
orbital inclination, $i\rightarrow \pi/2$ (edge on).

If a binary pulsar system with $i\rightarrow \pi/2$  still has
significant $\dot{P}_b$ (1PN),  then  Eq($\ref{wkdsme2}$)
corresponding to  Wex $\&$ Kopeikin (1999) is not supported, other
wise  Eq($\ref{wkdsme1}$) corresponding to this paper is not
supported.

The discrepancy between  Eq($\ref{wkdsme1}$) and
Eq($\ref{wkdsme2}$) is due to the discrepancy on $\dot{\omega}$,
which is caused by different degrees of freedom used in   this
paper and Wex $\&$ Kopeikin (1999).

Eq($\ref{wkdsme1}$) and Eq($\ref{wkdsme2}$) have important
property in common, that is $\dot{\omega}^S$ (and therefore,
$\dot{\omega}^{obs}$) is obtained by the transformation from the
J-coordinate system to the observer's coordinate system, in which
all the three triads are  at rest to  SSB.

\subsection{Discrepancy between Damour $\&$ Sch\"afer and Wex $\&$
Kopeikin }

Damour $\&$ Sch\"afer (1988) express the orbital precession
velocity as,
\begin{equation} \label{ds1}
\dot{\bf \Omega}_S=\frac{d\Omega_S}{dt}{\bf
K}_0+\frac{d\omega}{dt}{\bf k}+\frac{d i}{dt}{\bf i}
 \ ,
\end{equation}
where ${\bf K}_0$  unit vector along is line of sight, which
defines the third vector of a reference triad (${\bf I}_0$, ${\bf
J}_0$, ${\bf K}_0$), where ${\bf I}_0$-${\bf J}_0$ corresponds to
plane of the sky. And the triad of the orbit is (${\bf i}$, ${\bf
j}$, ${\bf k}$), in which ${\bf k}$ corresponds to $\hat{\bf L}$.
${\bf i}$ is the nodal vector determined by the  intersection of
the two planes (notice that it is different from the scaler, $i$,
which represents the orbital inclination), as shown in Fig~2b. By
Eq($\ref{ds1}$), and the relations between the reference triad,
components of $\dot{\bf \Omega}_S$ are obtained (Damour $\&$
Sch\"afer 1988).
\begin{equation} \label{ds2}
\frac{d\omega}{dt}=\frac{1}{\sin^2 i}[\dot{\bf \Omega}_S\cdot{\bf
k}-\dot{\bf \Omega}_S\cdot{\bf K}_0\cos i]
 \ ,
\end{equation}
\begin{equation} \label{ds3}
\frac{d\Omega_S}{dt}=\frac{1}{\sin^2 i}[\dot{\bf
\Omega}_S\cdot{\bf K}_0-\dot{\bf \Omega}_S\cdot{\bf k}\cos i]
 \ ,
\end{equation}
\begin{equation} \label{ds4}
\frac{d i}{dt}=\dot{\bf \Omega}_S\cdot{\bf i}
 \ .
\end{equation}
The S-L coupling induced $\dot{\omega}$ given by Eq($\ref{ds2}$)
is $\dot{\omega}\sim \dot{\Omega}_S\sim $1.5PN. Therefore, Damour
$\&$ Sch\"afer (1988) predict insignificant $\dot{\omega}$
($\dot{\omega}^S$) which is much smaller than 1PN, and therefore,
the corresponding $\dot{P}_b$ is also insignificant.

Thus it seems strange that Damour $\&$ Sch\"afer (1988) and Wex
$\&$ Kopeikin (1999) start from same orbital precession velocity
given by Eq($\ref{we3}$), but predict different observational
effect.

This is because $\dot{\omega}^S$  of Wex $\&$ Kopeikin is
calculated in a coordinate system with axes ($x^{\prime}$,
$y^{\prime}$, $z^{\prime}$) at rest to  SSB as shown in Fig~2a.
Whereas, $\dot{\omega}$ ($\dot{\omega}^S$) of Damour $\&$
Sch\"afer is calculated in a coordinate system with triad (${\bf
K}_0$, ${\bf k}$, ${\bf i}$), which is not at rest to  SSB as
shown in Fig~2b. Obviously ${\bf i}$ (point A), which is the
intersection of the plane of the sky and the orbital plane of a
binary system, is not static in the coordinate system
($x^{\prime}$, $y^{\prime}$, $z^{\prime}$) of Wex $\&$ Kopeikin
(1999), so is ${\bf k}$. In other words, triad (${\bf K}_0$, ${\bf
k}$, ${\bf i}$) has non-zero acceleration relative to SSB,
therefore, effects calculated based on such triad cannot be
compared with observation directly.

Obviously, if Damour $\&$ Sch\"afer's $\dot{\omega}$ is also
calculated in the coordinate system as that of Wex $\&$ Kopeikin,
then Eq($\ref{ds2}$) reduces to Eq($\ref{wkdsme2}$). Thus, the
discrepancy between Damour $\&$ Sch\"afer (1988) and  Wex $\&$
Kopeikin (1999) is that the former calculated in a coordinate
systems which is not at rest to SSB; whereas, the latter is at
rest to SSB. On the other hand, the discrepancy between  Wex $\&$
Kopeikin (1999) and this paper is different degrees of freedom
used in the calculation of equation of motion of a binary system.
The relationship of the three kinds of S-L coupling effects is
shown in Fig~4.


\section{Confrontation with observation}
The precise timing measurement on two typical binary pulsars,
PSR~J2051-0827 and PSR~J1713+0747, provides evidence on  whether
$\dot{\omega}^S$ and $\dot{P}_b$ is 1PN or 1.5PN, but it still
difficult to distinguish which 1PN effect is valid,  Wex $\&$
Kopeikin (1999) or this paper.

The orbital motion causes a delay of $T={\bf r}_1\cdot{\bf
K}_0/c=r_1(t)\sin\omega^{\rm obs}(t)\sin i(t)/c$ in the pulse
arrival time, where ${\bf r}$ is the pulsar position vector and
${\bf K}_0$ is the unit vector of the line of sight. The residual
$\delta T={\bf r}_1\cdot{\bf K}_0/c-({\bf r}_1\cdot{\bf K}_0/c)_K$
of the time delay compared with the Keplerian value is of interest
(Lai et al. ~\cite{lai}). Averaging over one orbit $r\approx a$,
and in the case $t\ll 1/|\dot{\omega}^{\rm obs}|$, the S-L
coupling induced residual is,
$$
\delta T=\frac{a_p}{c}\cos i\frac{di}{dt}t\sin\omega^{\rm
obs}+\frac{\dot{a}_p\sin i}{c}t\sin\omega^{\rm obs}
  +\frac{a_p\sin
i}{c}\dot{\omega}^{\rm obs}t\cos\omega^{\rm obs}
$$
\begin{equation}
\label{conf1} =\dot{x}_1t\sin\omega^{\rm obs}+
\frac{\dot{a}}{a}xt\sin\omega^{\rm obs}+\dot{\omega}^{\rm
obs}xt\cos\omega^{\rm obs}
 \ .
\end{equation}
By Eq.($\ref{dadt}$), we have ${\dot{a}}/{a}\sim J/a^3\sim
\dot{\omega}^{\rm obs}$, therefore, the second and third term at
the right hand side of Eq.($\ref{conf1}$) cannot be distinguished
in the current treatment of pulsar timing. In other words, the
effect of $\dot{a}$ can be absorbed by $\dot{\omega}^{\rm obs}$.

\subsection{PSR~J2051-0827}
As discussed above, the second term at the right hand side of
Eq.($\ref{e3c}$) can be absorbed by $\dot{\omega}^{\rm obs}$,
therefore $\dot{x}\approx\dot{x}_1$, and by Eq.($\ref{dotx1}$), we
have
\begin{equation}
\label{obs1} \dot{\Omega}=-\frac{\dot{x}}{x}\frac{\tan
i}{\sin\lambda_{LJ}\sin\Omega_0}=-\frac{di}{dt}\frac{1}{\sin\lambda_{LJ}\sin\Omega_0}
\ \ .
\end{equation}
According to optical observations, the system is likely to be
moderately inclined with an inclination angle $i\sim 40^{\circ}$
(Stappers et al. 2001).
By the measured results of $x=0.045\,$s, $\dot{x}=-23(3)\times
10^{-14}$ (Doroshenko et al. 2001), and by assuming
$\sin\lambda_{LJ}\sin\Omega_0=2\times 10^{-3}$, Eq.($\ref{obs1}$)
can be written  in magnitude,
\begin{equation}
\label{2051a}
 \dot{\Omega}=(\frac{\dot{x}_{}}{2.3\times 10^{-13}
})(\frac{x}{0.045})^{-1}(\frac{\tan i_{}}{\tan
40^{\circ}})(\frac{\sin\lambda_{LJ}\sin\Omega_0}{2\times10^{-3}})^{-1}
 \sim 2\times 10^{-9}\,(\rm s^{-1}) \ \ .
\end{equation}
In the following estimation of this section all values are
absolute values.
By Eq.($\ref{STomega}$), we can assume
$\ddot{\omega}^{S}\sim(\dot{\omega}^{S})^2\sim
\dot{\Omega}^2\approx 4\times 10^{-18}$. Usually
$\ddot{\omega}^{S}$ can vary in a large range, i.e.,
$\ddot{\omega}^{S}>(\dot{\omega}^{S})^2$, depending on different
combination of parameters, such as  binary parameters,  magnitude
and orientation of $S_1$ and $S_2$. In this paper we assume that
$\ddot{\omega}^{S}\sim(\dot{\omega}^{S})^2$. Thus from
Eq.($\ref{ez2}$) we have,
\begin{equation}
\label{2051b}\dot{P}_{b}=\frac{1}{2\pi}(\frac{\ddot{\omega}^{S}}{4\times
10^{-18}})(\frac{{P}_{b}}{0.099\,{\rm d}})^2\sim 5\times 10^{-11}
({\rm s\,s^{-1}}) \ \ .
\end{equation}
By Eq.($\ref{ef10ad}$), we can estimate $d^3{\omega}^{S}/dt^3\sim
{\dot{\Omega}}^3\approx 8\times 10^{-27}\,$s$^{-3}$, similarly, we
can estimate $d^4\omega^{S}/dt^4\sim \dot{\Omega}^4\approx
16\times 10^{-36}\,$s$^{-4}$.

Therefore, by Eq.($\ref{ez2a}$) and Eq.($\ref{ez2b}$) we have
$\ddot{P}_{b}\sim 9\times 10^{-20}\,$s$^{-1}$ and
$d^3{P}_{b}/dt^3\sim 2\times 10^{-28}\,$s$^{-2}$. By
Eq.($\ref{e3c}$) and Eq.($\ref{e3cc}$), $\ddot{x}/\dot{x}\sim
\dot{\Omega}\sim 2\times 10^{-9}$s$^{-1}$, which is consistent
with observation as shown in Table~2.

Therefore, once $\dot{x}$ is in agreement with the observation,
the corresponding $\dot{\omega}^{S}$ can make the derivatives of
$P_b$ be consistent with observation as shown in Table~2. Whereas,
the effect derived from Damour $\&$ Sch\"afer's equation cannot
explain the significant derivatives of ${P}_b$.
\subsection{PSR~J1713+0747}
By the measured parameters, $x=32.3\,$s, $|\dot{x}|=5(12)\times
10^{-15}$, $i=70^{\circ}$ (Camilo et al. 1994), and by assuming
$\sin\lambda_{LJ}\sin\Omega_0=1\times 10^{-4}$, then in magnitude
we have,
\begin{equation}
\label{1713a}
 \dot{\Omega}=(\frac{\dot{x}_{}}{5\times 10^{-15}
})(\frac{x}{32.3})^{-1}(\frac{\tan i_{}}{\tan
70^{\circ}})(\frac{\sin\lambda_{LJ}\sin\Omega_0}{1\times10^{-4}})^{-1}
\sim 4\times 10^{-12} ({\rm s^{-1}}) \ \ ,
\end{equation}
similarly we have,
\begin{equation}
\label{1713b}\dot{P}_{b}=\frac{1}{2\pi}(\frac{\ddot{\omega}^{S}}{16\times
10^{-26}})(\frac{{P}_{b}}{67.8\,{\rm d}})^2\sim 1\times 10^{-10}
({\rm s\,s^{-1}}) \ \ .
\end{equation}
The comparison of observational and predicted  variabilities are
shown Table~3, which are well consistent. Notice that
$\dot{x}^{\rm obs}$ and $\dot{P}_b^{\rm obs}$ measured in these
two typical binary pulsars cannot be interpreted by the
gravitational radiation induced $\dot{x}$ and $\dot{P}_b$, since
they are 3 or 4 order of magnitude smaller than those of the
observational ones.

\subsection{PSRs~J0737-3039A and B}
PSRs~J0737-3039A and B is a double-pulsar system with
$P_b=0.102251563(1)$day, advance of periastron,
$\dot{\omega}=16.90(1)$deg/yr and orbital inclination angle
$i=87.7^{+17}_{-29}$ (Burgay et al. 2003, Lyne et al. 2004). This
binary pulsar system with $i\approx\pi/2$ may tell us not only
whether $\dot{\omega}^S$ and $\dot{P}_b$ is significant or not,
but also which significant $\dot{\omega}^S$ and $\dot{P}_b$
effects (Wex $\&$ Kopeikin or this paper) is valid.

As given by Eq.($\ref{ef10}$) the measured $\dot{\omega}$ is the
sum of relativistic advance of periastron and the S-L coupling
induced advance of periastron,
\begin{equation}
\label{0737a} \dot{\omega}^{\rm
obs}=\dot{\omega}^{GR}+\dot{\omega}^{S}=16.90 \ (deg/yr)
 \ \,.
\end{equation}
$\dot{\omega}^{GR}$ and $\dot{\omega}^{S}$ are both 1PN. In order
of magnitude one can estimate
\begin{equation}
\label{0737b} \dot{\omega}^{S}\sim\dot{\omega}^{\rm obs}=16.90
 \ (deg/yr)
 \ \,.
\end{equation}
Thus we have $\ddot{\omega}^S\sim(\dot{\omega}^{S})^2\approx
8.8\times10^{-17}$s$^{-2}$.  By the same treatment of
PSR~J2051-0827 and PSR~J1713+0747 we have,
\begin{equation}
\label{0737c}\dot{P}_{b}=\frac{1}{2\pi}(\frac{\ddot{\omega}^{S}}{8.8\times
10^{-17}})(\frac{{P}_{b}}{0.1\,{\rm d}})^2\sim 1\times 10^{-9}
({\rm s\,s^{-1}}) \ \ .
\end{equation}
The treatment $\dot{\omega}^{S}\sim\dot{\omega}^{\rm obs}$ and
$\ddot{\omega}^S\sim(\dot{\omega}^{S})^2$ might over estimate the
S-L coupling induced $\dot{P}_{b}$ for one or even two order of
magnitude. Nevertheless, the S-L coupling induced $\dot{P}_{b}$ is
likely much larger than that caused by the gravitational
radiation, $\dot{P}_b^{GR}=-1.2\times 10^{-12}$ ${\rm s\,s^{-1}}$
(Burgay et al. 2003).  Since the observational $\dot{P}_b$ will be
given soon(Burgay et al. 2003), whether $\dot{P}_b$ is significant
or not can be tested on this binary pulsar system.

The specific orbital inclination of this binary pulsar system can
tell us more about the S-L coupling induced effects. By
$i=87.7^{+17}_{-29}$ of PSRs~J0737-3039A and B, we have $\cot
i\approx 0.04$, therefore Wex and Kopeikin's $\dot{\omega}^{S}$
given by Eq.($\ref{wkdsme2}$) should be much smaller than that of
this paper's $\dot{\omega}^{S}$ given by Eq.($\ref{wkdsme1}$).

Thus we have two ways to test the validity the S-L coupling effect
given by  Wex and Kopeikin (1999) and this paper.

The first one is that by
$\ddot{\omega}^S\sim(\dot{\omega}^S)^2\propto (\cot i)^2$, and
$\dot{P}_{b}\propto (\dot{\omega}^S)^2$, we have
$(\dot{P}_{b})_{WK}\sim (\cot i)^2\dot{P}_{b}$. Thus the magnitude
of $\dot{P}_{b}$ corresponding to Wex and Kopeikin should not
exceed $1.6\times10^{-12}$ ${\rm s\,s^{-1}}$, which is close to
gravitational wave induced $\dot{P}_{b}$.

In other words if the measured magnitude of  $\dot{P}_{b}$ of
PSRs~J0737-3039A and B is not much larger than that gravitational
wave induced $\dot{P}_{b}$, then the expression of this paper can
be excluded.

The second one is based on  Eq.($\ref{ez2}$) to Eq.($\ref{ez2b}$),
from which we have,
\begin{equation}
\label{chac} \frac{|\ddot{P}_{b}|}{|d^3{P}_{b}/dt^3|}\sim
\frac{|\dot{P}_{b}|}{|\ddot{P}_{b}|}\sim
\frac{1}{|\dot{\omega}^{S}|}
\end{equation}
Since $\dot{\omega}^{S}$ corresponds to Wex and Kopeikin is nearly
two order of magnitude smaller than the relativistic advance of
periastron, $\dot{\omega}^{GR}$;  whereas, $\dot{\omega}^{S}$ of
this paper is same order of magnitude of $\dot{\omega}^{GR}$.
Therefore, if the ratio of Eq.($\ref{chac}$) is close to
$1/\dot{\omega}^{GR}$ then this paper is supported.

The precise measurement of derivatives of ${P}_b$ of this binary
pulsar system may provide chance to test which one is favored.


\section{Discussion and conclusion}
BO and AK's orbital precession velocity has been treated as
equivalent, since BO and AK give equivalent torque, $\dot{\bf L}$,
as shown in Eq.($\ref{e1.13}$) and Eq.($\ref{spin3}$)
respectively. However Eq.($\ref{e1.13}$) and Eq.($\ref{spin3}$)
actually correspond to two different orbital precession
velocities, Eq.($\ref{we3}$) and Eq.($\ref{e1aa4}$) respectively,
this is because the same torque can cause different effect when a
dynamic system is calculated under different degrees of freedom.
BO's orbital precession velocity is actually obtained under 10
degrees of freedom; whereas, AK's is under 8 degrees of freedom.
Correspondingly the former one violates the triangle constraint
and the latter one satisfies it.

The discrepancy in physics results discrepancy in observation.
Eq.($\ref{we3}$) and Eq.($\ref{e1aa4}$) correspond to different
combinations of $\Omega$ and $\omega$ ($\Omega$ and $\omega$ are
defined in Fig.2), and since observational effect depends on
$\Omega$ and $\omega$ instead of $\dot{\bf L}$, therefore, the
equivalent value in $\dot{\bf L}$ may correspond to different
observational effects. Concretely, BO and AK give same $\Omega$,
but different $\omega$ (notice that $\Omega$ and $\omega$ are
components of the vectors given by Eq.($\ref{we3}$) or
Eq.($\ref{e1aa4}$)). And by Eq.($\ref{ef9a}$) and
Eq.($\ref{ef9}$), the observed advance of precession of periastron
depends on both $\Omega$ and $\omega$, thus BO and AK must
correspond to different observational effects.

In the calculation of  Sect~4 and Sect~5, we can see the influence
of the degree of freedom and physical constraint on the results of
equation of motion, perturbation and observational effects.

The S-L coupling induced precession of orbit can  cause an
additional time delay to the time of arrivals (TOAs), which can be
absorbed by the orbital period. And since the additional time
delay itself is a function of time, therefore orbital period
change, $\dot{P}_b$, appears. Actually  $\dot{P}_b$ corresponds to
$\ddot{\omega}^S$ as shown in Eq.($\ref{ez2}$), which cannot be
absorbed by $\dot{\omega}^{GR}$ ($\dot{\omega}^S$ can be absorbed
by $\dot{\omega}^{GR}$). Therefore, the higher order of
derivatives of orbital period provide good chance to test
different models. The observation of $\dot{P}_b$, $\ddot{P}_b$ and
$d^3{P}_b/dt^3$ on PSR J~2051-0827  supports significant S-L
coupling induced effects.

This paper for the first time points out that Wex $\&$ Kopeikin's
expression in 1999 actually corresponds to
 significant $\dot{\omega}^S$ and $\dot{P}_b$, however
it is not equivalent to the significant $\dot{\omega}^S$ and
$\dot{P}_b$ given by this paper. Precise measurement of
$\dot{P}_b$, $\ddot{P}_b$ and $d^3{P}_b/dt^3$ of specific binary
pulsars with orbital inclination $i\rightarrow \pi/2$, like
PSRs~0737-3039A and B, may provide chance to test the validity of
the results corresponding to Wex $\&$ Kopeikin (1999) and that of
this paper.

\acknowledgments{
 I thank T.~Huang for help in clarifying the theoretical part of
this paper. I thank R.N.~Manchester for his help in understanding
pulsar timing measurement. I thank T. Lu for useful comments
during this work. I thank W.T.~Ni and C.M.~Xu for useful
suggestions in the presentation of this paper. I thank E.K.~Hu,
A.~R\"udiger, K.S.~Cheng, N.S.~Zhong, and Z.G.~Dai for continuous
encouragement and help.  I also thank Y.~Li, Z.X.~Yu, C.M.~Zhang,
L.~Zhang, Z.~Li, H.~Zhang, S.Y.~Liu, X.N.~Lou, X.S.~Wan for useful
discussions. }

\section{Appendix}
By $\widetilde{S}={\bf a}_{so}\cdot \hat{\bf n}$ and  $T={\bf
a}_{so}\cdot \hat{\bf t}$, $\frac{d\Omega}{dt}$ and
$\frac{d\omega}{dt}$ have been given by Eq.($\ref{Womega}$),
Eq.($\ref{Womegaav}$), Eq.($\ref{STomega0}$), and
Eq.($\ref{STomega}$), following the standard procedure for
computing perturbations of orbital elements Roy~(\cite{roy}).
Similarly, four other elements can be given:
\begin{equation}
\label{dadt}
\frac{da}{dt}=\frac{2}{n\sqrt{1-e^2}}(\widetilde{S}e\sin
f+\frac{p\widetilde{T}}{r})\ \,,
\end{equation}
\begin{equation}
\label{dadtav}
<\frac{da}{dt}>=\frac{4\sigma_1J(1+e^2)}{(1-e^2)^{5/2}a^2}(P_xQ_y-P_yQ_x)
\ \,,
\end{equation}

\begin{equation}
\label{dedt}
\frac{de}{dt}=\frac{\sqrt{1-e^2}}{na}[\widetilde{S}\sin
f+\widetilde{T}(\cos E+\cos f)]\ \,,
\end{equation}
\begin{equation}
\label{dedtav}
<\frac{de}{dt}>=\frac{4\sigma_1Je}{(1-e^2)^{3/2}a^3}(P_xQ_y-P_yQ_x)
\ \,,
\end{equation}

\begin{equation} \label{didt}
\frac{d \lambda_{LJ}}{dt}=\frac{\widetilde{W}r\cos
(\omega+f)}{na^2\sqrt{1-e^2}}\frac{1}{\sin\lambda_{LJ}} \ ,
\end{equation}
$$
<\frac{d \lambda_{LJ}}{dt}>=\frac{3\cos\lambda_{LJ}
}{2a^3(1-e^2)^{3/2}\sin\lambda_{LJ}}(P_z\cos\omega+Q_x\sin\omega)
[(P_yQ_z-P_zQ_y)(S_x\sigma_1+S_{2x}\sigma_2)$$
\begin{equation} \label{Womegaavi}
+(P_zQ_x-P_xQ_z)(S_y\sigma_1+S_{2y}\sigma_2)
+(P_xQ_y-P_yQ_x)(S_z\sigma_1+S_{2z}\sigma_2)] \ ,
\end{equation}

\begin{equation} \label{dmdt}
\frac{d
\epsilon}{dt}=\frac{e^2}{1+\sqrt{1-e^2}}\frac{d\varpi}{dt}+
2\frac{d\Omega}{dt}(1-e^2)^{1/2}(\sin^2\frac{\lambda_{LJ}}{2})-\frac{2r\widetilde{S}}{na^2}
 \ ,
\end{equation}
where
$\frac{d\varpi}{dt}=\frac{d\omega^{\prime}}{dt}+2\frac{d\Omega}{dt}(\sin^2\frac{\lambda_{LJ}}{2})$.
$$
<\frac{d
\epsilon}{dt}>=\frac{e^2}{1+\sqrt{1-e^2}}<\frac{d\varpi}{dt}>
+
2<\frac{d\Omega}{dt}>(1-e^2)^{1/2}(\sin^2\frac{\lambda_{LJ}}{2})
$$
\begin{equation} \label{dmdtav}
-\frac{4\sigma_1J}{a^3(1-e^2)}(P_xQ_y-P_yQ_x)
 \ .
\end{equation}

\begin{table}
\begin{center}
\caption{Comparison of S-L coupling induced  variabilities given
by different authors}
\begin{tabular}{crrrr}
\hline \hline

& DS (1988) & WK (1999)   & This paper  & evidence\\\hline

$\dot{\Omega}$  in J-co &  & 1PN & 1PN &  \\

$\dot{\omega}$  in J-co & & 1PN & 1.5PN &  \\

$\dot{\omega}^{\rm S}$ & $\sim\dot{\Omega}_S$ (of
Eq.($\ref{we3}$)) & $-\dot{\lambda}_{LJ} \cot i\sin\Omega$ &
$\dot{\Omega}-\dot{\lambda}_{LJ} \cot i\sin\Omega$
&   \\

$\dot{\omega}^{\rm obs}$ &  $\dot{\omega}^{GR}$+1.5PN &
$\dot{\omega}^{GR}$+1PN
& $\dot{\omega}^{GR}$+1PN  \\

$\dot{P}_b$ & $\dot{P}_{b}^{GR}$ & $|\dot{P}_{b}^{\rm obs}|\gg
|\dot{P}_{b}^{GR}|$ & $|\dot{P}_{b}^{\rm obs}|\gg
|\dot{P}_{b}^{GR}|$ &  $|\dot{P}_{b}^{\rm obs}|\gg |\dot{P}_{b}^{GR}|$ \\

when $i\rightarrow \pi/2$ & &
$\dot{P}_b^{obs}\rightarrow\dot{P}_b^{GR}$ &
$|\dot{P}_b^{obs}|\gg |\dot{P}_b^{GR}|$ & \\

\hline \hline
\end{tabular}
\end{center}
{\small   J-co represents J-coordinate system, $\dot{P}_b^{GR}$
represents orbital period change due to gravitational radiation
predicted by General Relativity. $\dot{\Omega}_S$ is given by
Eq.($\ref{we3}$) which is 1.5PN, whereas, $\dot{\Omega}$ and
$\dot{\lambda}_{LJ}$ are 1PN, as given by Eq.($\ref{Womegaav}$)
and  Eq.($\ref{didt}$) respectively.
}
\end{table}


\begin{table}
\begin{center}
\caption{Measured parameters compared with the geodetic precession
induced ones in PSR~J2051$-$0827}
\begin{tabular}{ll}
\hline \hline
 observation & WK $\&$ this paper     \\ \hline
 $\dot{x}^{\rm obs}
\approx-23(3)\times 10^{-14}$
  & $\dot{x}=\dot{x}^{\rm obs}$ \ \    \\

$({\ddot{x}}/{\dot{x}})^{\rm obs}
 \leq-3.0\times 10^{-9}$s$^{-1}$
  & $|{\ddot{x}}/{\dot{x}}|\approx 2\times 10^{-9}$s$^{-1}$    \\

$\dot{P}_{b}^{\rm obs}=-15.5(8)\times 10^{-12}$ & $|\dot{P}_{b}|=
|\frac{\ddot{\omega}^{S}P_{b}^{2}}{2\pi}| \sim 5\times
10^{-11}$  \\

$\ddot{P}_{b}^{\rm obs}=2.1(3)\times 10^{-20}$s$^{-1}$ &
$|\ddot{P}_{b}|=
|\frac{P_b^2}{2\pi }\frac{d^3{\omega}^{S}}{dt^3}|
\sim 9\times 10^{-20}$s$^{-1}$  \\

$\frac{d^3{P}_{b}^{\rm obs}}{dt^3}=3.6(6)\times 10^{-28}$s$^{-2}$
& $|\frac{d^3{P}_{b}}{dt^3}|=
|\frac{P_b^2}{2\pi}\frac{d^4{\omega}^{S}}{dt^4}|
 \sim 2\times 10^{-28}$s$^{-2}$
\\ \hline \hline
\end{tabular}
\end{center}
{
 }
\end{table}


\begin{table}
\begin{center}
\caption{Measured parameters compared with the geodetic precession
induced ones in PSR~J1713+0747}
\begin{tabular}{ll}
\hline \hline
 observation &  WK $\&$ this paper    \\ \hline
 $|\dot{x}|^{\rm obs}
=5(12)\times 10^{-15}$
  & $\dot{x}=\dot{x}^{\rm obs}$    \\

$\dot{P}_{b}^{\rm obs}=1(29)\times 10^{-11}$ &
$|\dot{P}_{b}|=|\frac{\ddot{\omega}^{S}P_{b}^{2}}{2\pi}|\sim
1\times 10^{-10}$
\\ \hline \hline
\end{tabular}
\end{center}
{\small
 }
\end{table}



\begin{figure}[t]
\begin{center}
\includegraphics[87,87][700,700]{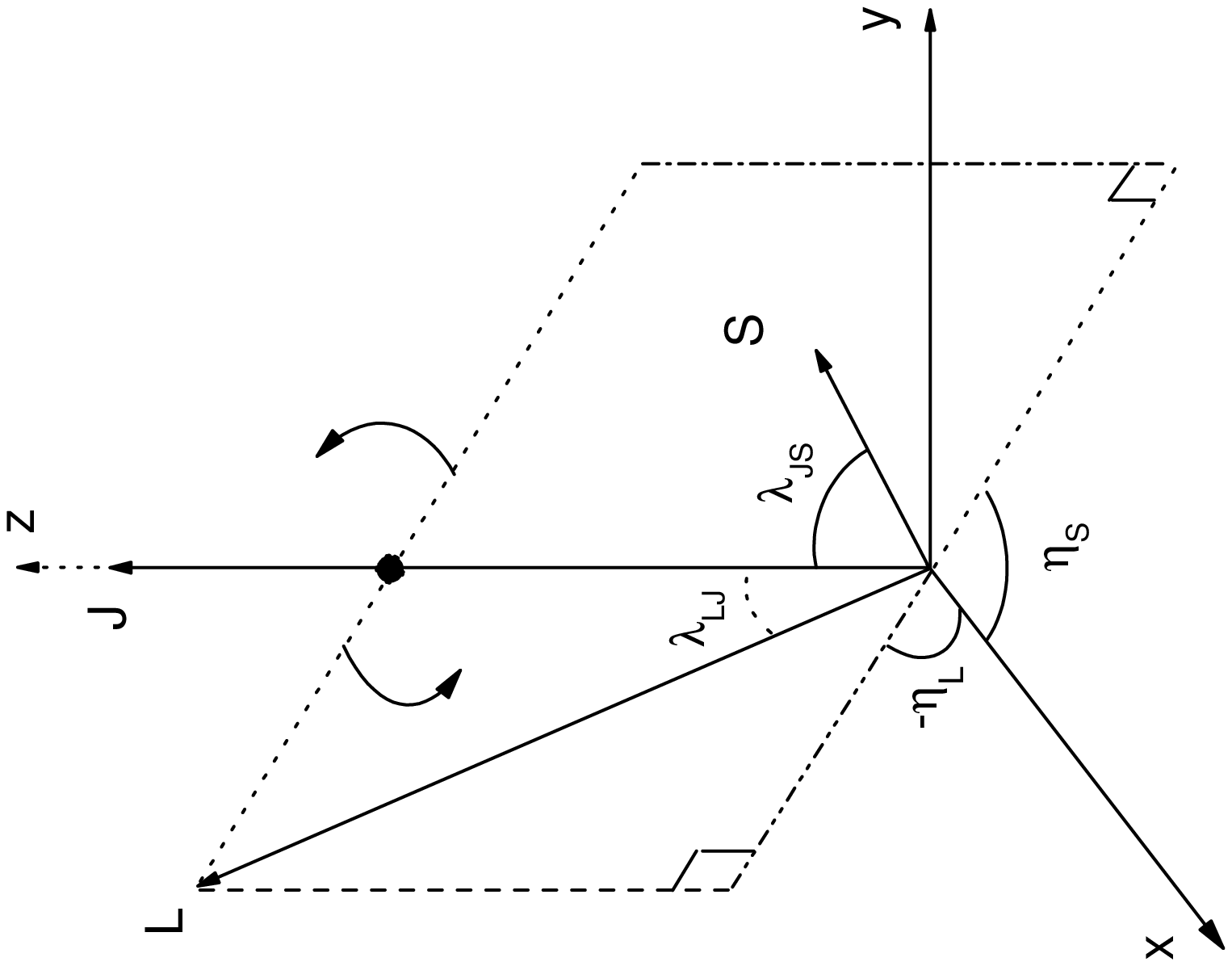}
\end{center}
\caption{In 1PN, the scenario of motion of a binary pulsar system
can be described as the rotation of the plane determined by ${\bf
L}$, ${\bf S}$ around the  fixed axis, ${\bf J}$. $\eta_L$  and
$\eta_S$ are precession phases of ${\bf L}$ and ${\bf S}$ in the
J-coordinate system respectively. }
\end{figure}

\begin{figure}[t]
\begin{center}
\includegraphics[87,87][700,700]{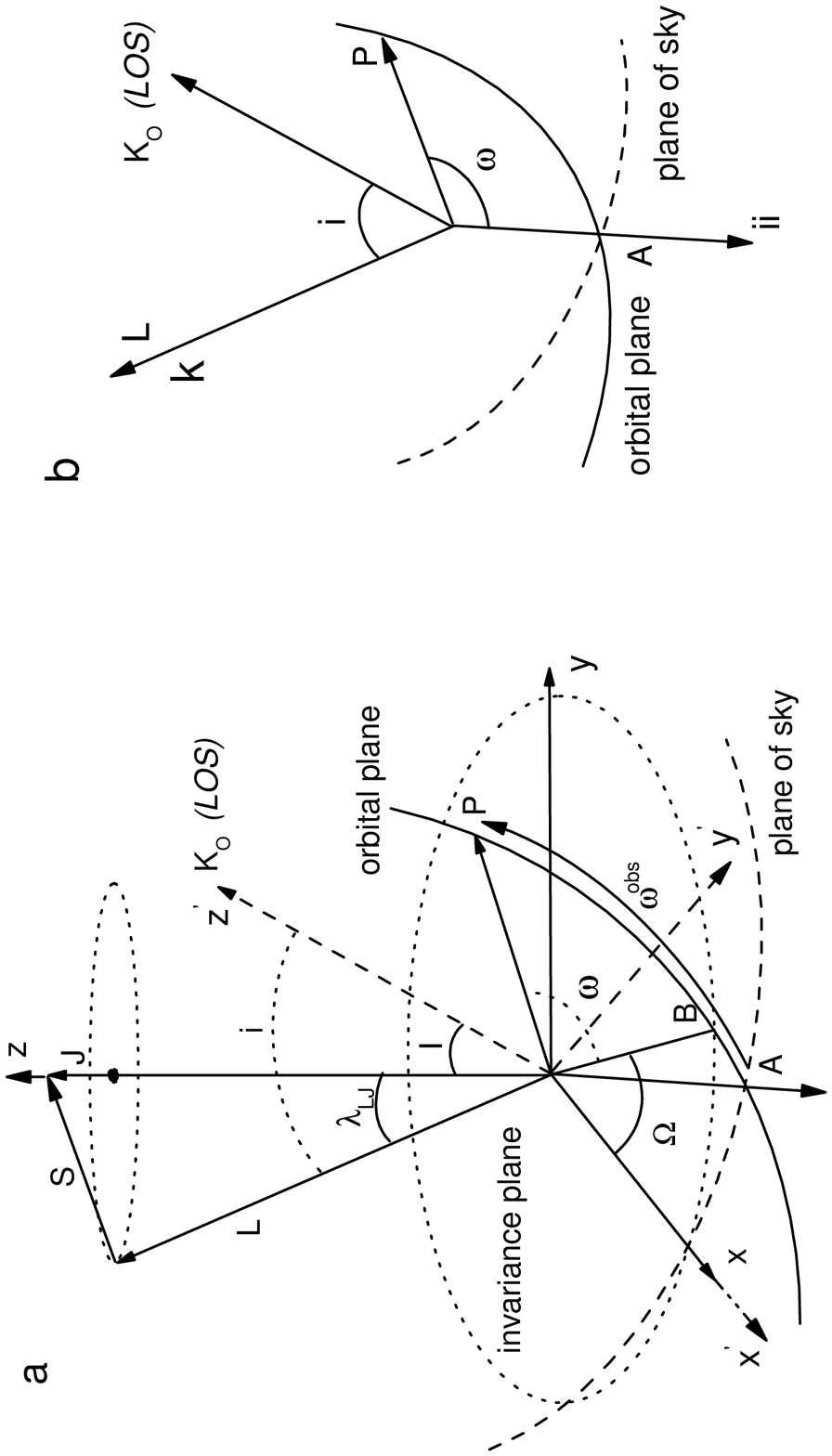}
\end{center}
\caption{Fig~2a: binary geometry and definitions of angles of Wex
$\&$ Kopeikin (1999) and this paper. The invariable plane (x-y),
as represented by the dotted ellipse, is perpendicular to the
total angular momentum, ${\bf J}$. The inclination of the orbital
plane with respect to the invariable plane is $\lambda_{LJ}$,
which is also the precession cone angle of  ${\bf L}$ around ${\bf
J}$. The orbital inclination with respect to the line of sight is
$i$. $\Omega$ is the longitude of the ascending node.  $\omega$ is
the longitude of the periastron from point B, and
$\dot{\omega}^{\rm obs}$ is longitude of the periastron from point
A. The J-coordinate system is determined by ($x$, $y$, $z$), and
the observer's coordinate system is determined by ($x^{\prime}$,
$y^{\prime}$, $z^{\prime}$). Fig~2b: coordinate system of Damour
$\&$ Sch\"afer (1988), which is determined by triad ($\hat{\bf
L}$, ${\bf i}$, ${\bf K}_0$). ii in Fig~2b represents ${\bf i}$. }
\end{figure}

\begin{figure}[t]

\begin{center}
\includegraphics[87,87][700,700]{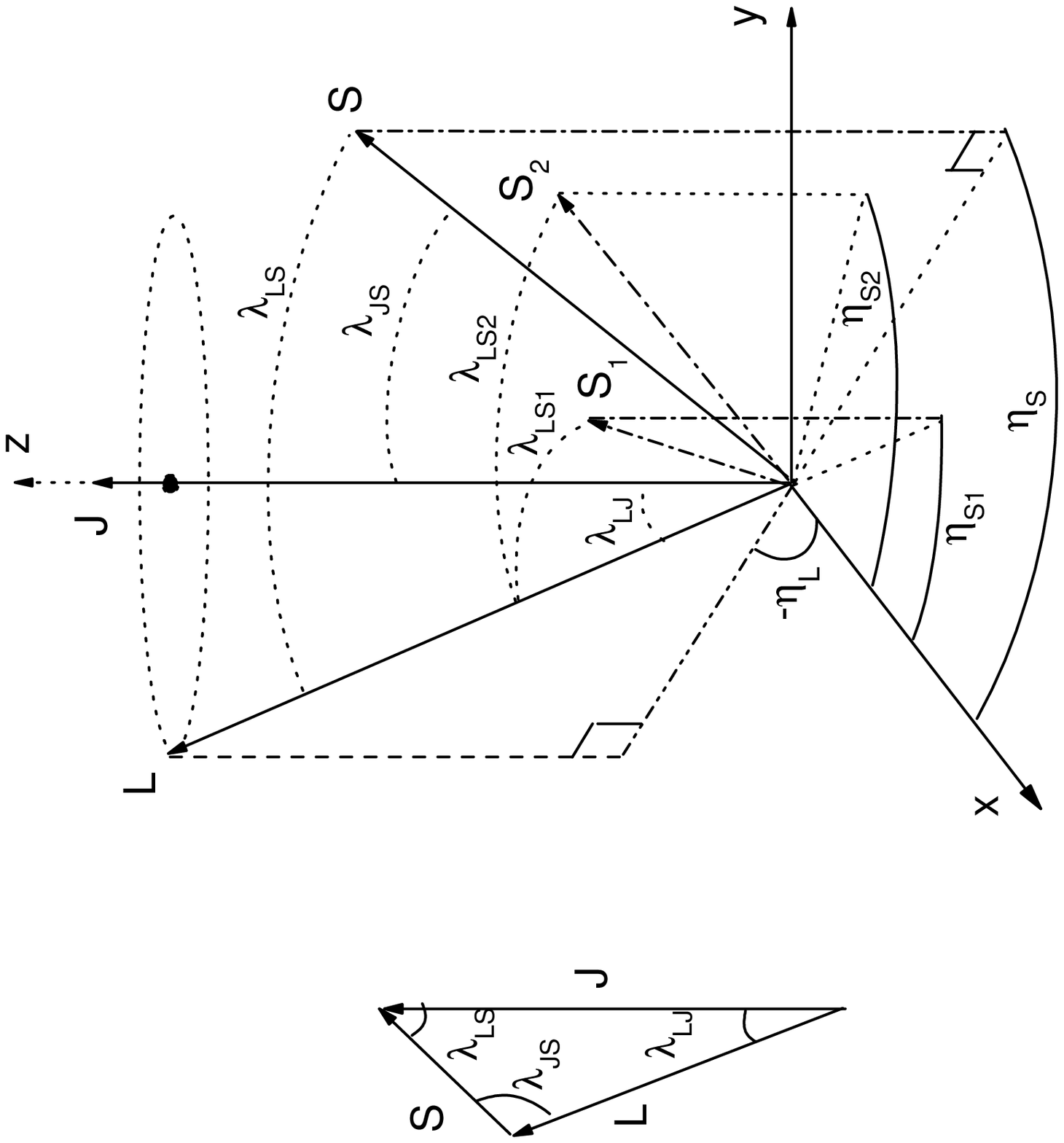}

\end{center}
\caption{Angles and orientation conventions relating
 vectors ${\bf
S}$, ${\bf L}$ and ${\bf J}$ to the coordinate system. x-y is the
invariance plane, $\eta_{S1}$,  $\eta_{S2}$, $\eta_S$ and $\eta_L$
are precession phases of ${\bf S}_1$, ${\bf S}_2$, ${\bf S}$ and
${\bf L}$, respectively. }
\end{figure}

\begin{figure}[t]

\begin{center}
\includegraphics[87,87][700,700]{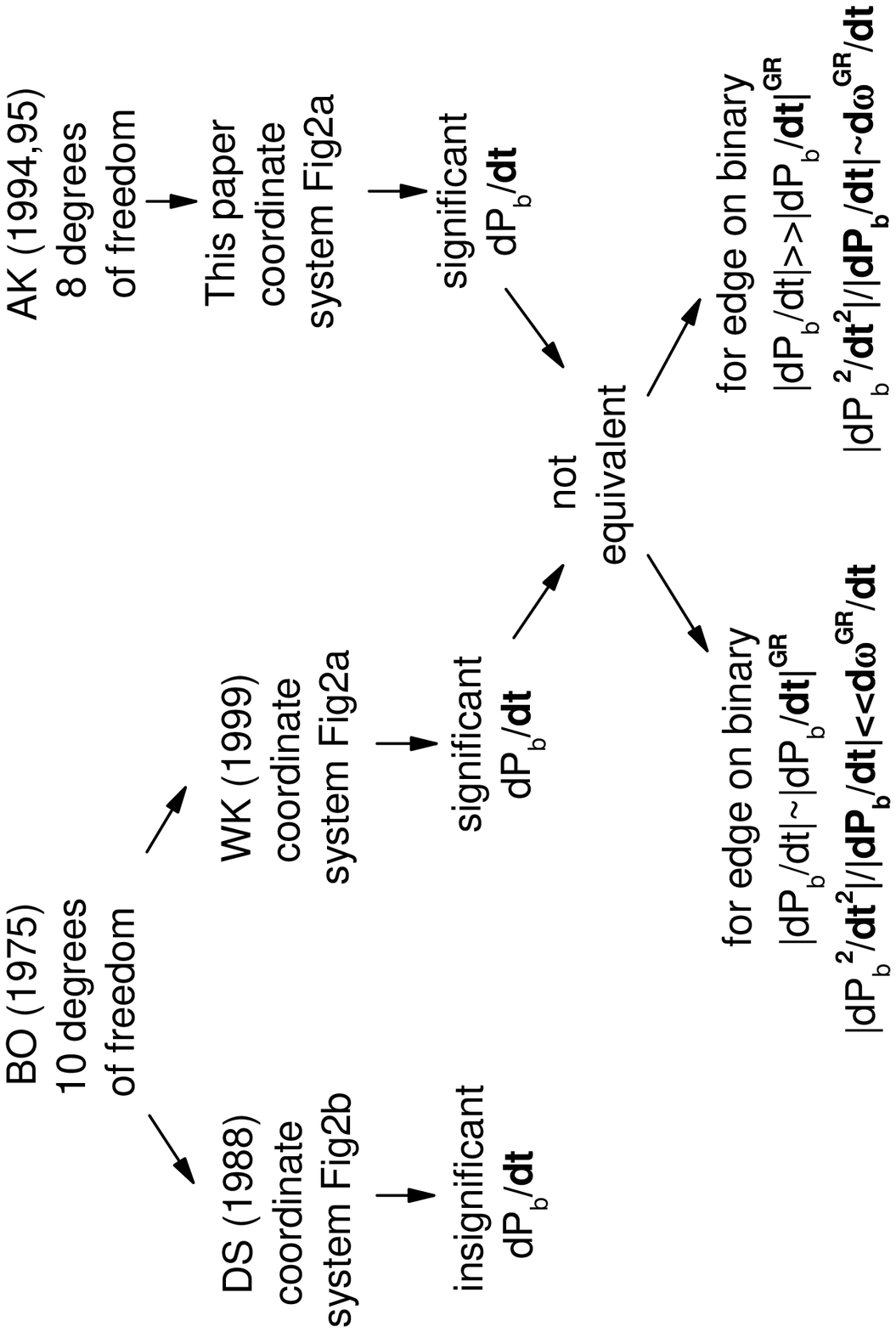}

\end{center}
\caption{The relationship of the three kinds of S-L coupling
induced orbital effects.
 }
\end{figure}

\end{document}